\documentclass{aa}  
\usepackage{graphicx}
\usepackage{txfonts}
\usepackage[colorlinks=true,citecolor=blue,linkcolor=blue,urlcolor=blue]{hyperref}
\usepackage{longtable}
\usepackage{multirow}
\usepackage{upgreek}
\def\ha{H$\upalpha$}

\begin{document}

   \title{A first look at spatially resolved star formation at $4.8<z<6.5$ with JWST FRESCO NIRCam slitless spectroscopy}
   
   \titlerunning{Spatially resolved star formation at $4.8<z<6.5$}
   
   \authorrunning{Matharu et al.}


    \author{Jasleen Matharu
           \inst{1}\fnmsep
           \inst{2},
           Erica J. Nelson\inst{3},
           Gabriel Brammer
           \inst{1}\fnmsep
           \inst{2},
           Pascal A. Oesch
           \inst{1}\fnmsep
           \inst{2}\fnmsep
           \inst{4},
           Natalie Allen
           \inst{1}\fnmsep
           \inst{2},
           Irene Shivaei
           \inst{5},
           Rohan P. Naidu
           \inst{6},
           John Chisholm
           \inst{7},
           Alba Covelo-Paz
           \inst{4},
           Yoshinobu Fudamoto
           \inst{8},
           Emma Giovinazzo
           \inst{4},
           Thomas Herard-Demanche
           \inst{9},
           Josephine Kerutt
           \inst{10},
           Ivan Kramarenko
           \inst{11},
           Danilo Marchesini
           \inst{12},
           Romain A. Meyer
           \inst{4},
           Gonzalo Prieto-Lyon
           \inst{1}\fnmsep
           \inst{2},
           Naveen Reddy
           \inst{13},
           Marko Shuntov
           \inst{1}\fnmsep
           \inst{2},
           Andrea Weibel
           \inst{4},
           Stijn Wuyts
           \inst{14} and
           Mengyuan Xiao
           \inst{4}}

   \institute{{Cosmic Dawn Center, Copenhagen, Denmark}
        \and
        {Niels Bohr Insitute, University of Copenhagen, Jagtvej 128, 2200 Copenhagen, Denmark}\\
        \email{jasleen.matharu@nbi.ku.dk}
        \and
        {Department for Astrophysical and Planetary Science, University of Colorado, Boulder, CO 80309, USA}
        \and    
        {Department of Astronomy, University of Geneva, Chemin Pegasi 51, 1290 Versoix, Switzerland}
        \and
        {Centro de Astrobiolog\'{i}a (CAB), CSIC-INTA, Carretera de Ajalvir km 4, Torrej\'{o}n de Ardoz, 28850, Madrid, Spain}
        \and
        {MIT Kavli Institute for Astrophysics and Space Research, 77 Massachusetts Avenue, Cambridge, 02139, Massachusetts, USA}
        \and
        {Department of Astronomy, University of Texas at Austin, 2515 Speedway, Austin, Texas 78712, USA}
        \and
        {Center for Frontier Science, Chiba University, 1-33 Yayoi-cho, Inage-ku, Chiba 263-8522, Japan}
        \and
        {Leiden Observatory, Leiden University, NL-2300 RA Leiden, Netherlands}
        \and
        {Kapteyn Astronomical Institute, University of Groningen, P.O. Box 800, 9700 AV Groningen, The Netherlands}
        \and
        {Institute of Science and Technology Austria (ISTA), Am Campus 1, 3400 Klosterneuburg, Austria}
        \and
        {Department of Physics \& Astronomy, Tufts University, MA 02155, USA}
        \and
        {Department of Physics and Astronomy, University of California, Riverside, 900 University Avenue, Riverside, CA 92521, USA}
        \and
        {Department of Physics, University of Bath, Claverton Down, Bath BA2 7AY, UK}
        }

   \date{Received XXX; accepted XXX}

  \abstract
   {We present the first results of the spatial distribution of star formation in 454 star-forming galaxies just after the epoch of reionisation ($4.8<z<6.5$) using \ha~emission-line maps and F444W imaging that traces the stellar continuum from the JWST FRESCO NIRCam Slitless Spectroscopy Survey. The {\ha} equivalent width profiles of star-forming galaxies across the main sequence at $z\sim5.3$ with stellar masses $6.8\leqslant$log($M_{*}$/M$_{\odot}$)$<11.1$ increase with radius, which provides direct evidence for the inside-out growth of star-forming galaxies just after the epoch of reionisation. \texttt{GALFIT} was used to calculate half-light radii, $R_{\mathrm{eff}}$, and central surface densities within 1 kiloparsec, $\Sigma_{1\mathrm{kpc}}$ of \ha~and the continuum. At a fixed stellar mass of Log$(M_{*}/\mathrm{M}_{\odot})=9.5$, $\Sigma_{1\mathrm{kpc, H}\upalpha}$ is $1.04\pm0.05$ times higher than $\Sigma_{1\mathrm{kpc, C}}$,  $R_{\mathrm{eff, H}\upalpha}$ is $1.18\pm0.03$ times larger than $R_{\mathrm{eff, C}}$ and both $R_{\mathrm{eff}}$ measurements are smaller than 1 kiloparsec. These measurements suggest the rapid build-up of compact bulges via star formation just after the epoch of reionisation. By comparison to analogous work done at lower redshifts with {\it Hubble} Space Telescope WFC3 slitless spectroscopy as part of the 3D-HST ($z\sim1$) and CLEAR ($z\sim0.5$) surveys, we find that $R_{\mathrm{eff}}(z)$ evolves at the same pace for {\ha} and the continuum, but $\Sigma_{1\mathrm{kpc}}(z)$ evolves faster for {\ha} than the stellar continuum. As a function of the Hubble parameter,  
   $\frac{R_{\mathrm{eff, H}\upalpha}}{R_{\mathrm{eff, C}}}=1.1h(z)$ and $\frac{\Sigma_{1\mathrm{kpc,H}\upalpha}}{\Sigma
_{1\mathrm{kpc,C}}}=h(z)^{1.3}$. These parametrisations suggest that the inside-out growth of the disk starts to dominate the inside-out growth of the bulge towards lower redshifts. This is supported by the redshift evolution in the EW({\ha}) profiles from FRESCO, 3D-HST, and CLEAR at fixed stellar mass and when star-forming progenitors are traced, in which in EW({\ha}) rapidly increases with radius within the half-light radius at $z\sim5.3$, but  EW({\ha}) increases only significantly with radius in the outer disk at $z\sim0.5$.}


   \maketitle
%

\section{Introduction}

The assembly of galaxies is partly controlled by their dark matter haloes in a Lambda cold dark matter ($\Lambda$CDM) Universe. The radial distribution of stars is set by the angular momentum distribution and the gas accretion rate is set by the halo mass \citep{White&Rees, Fall1980, Dalcanton1997, VanDenBosch2001, Dekel2013}. The sizes of galaxies should therefore be proportional to the sizes of their dark matter haloes \citep{Mo1998}. The formation of stars in galaxies should progress towards larger galactocentric radii with time. Evidence for the inside-out growth of galaxies via star formation is required to confirm this picture. 

The earliest studies in the local Universe used narrow-band imaging targeting wavelengths that are sensitive to star formation on different timescales \citep{Hodge1983,Athanassoula1993,Ryder1994,Kenney1999,Koopmann2004a,Koopmann2004,Koopmann2006,Cortes2006,Crowl2006,MunozMateos2007,Abramson2011a,Vollmer2012,Gavazzi2013,Kenney2015,Abramson2016b,Lee2017,Gavazzi2018,Cramer2019,Boselli2020}. One of the most effective methods of mapping on-going star formation in galaxies at low and high redshift is to observe the \ha~emission. The ultraviolet radiation emitted by young O- and B-type stars ionises the hydrogen gas surrounding them. Recombination of the hydrogen atoms leads to emission in {\ha} \citep{Kennicutt1998}. Since these stars have lifetimes of $\sim10$~Myr, \ha~emission-line maps allow us to observe where star formation occurred in the galaxies over the past $\sim10$~Myr. Rest-frame optical imaging of galaxies provides a good tracer of the integrated star formation history, the so-called stellar continuum. By comparing the spatial distribution of {\ha} to the spatial distribution of the same galaxy in rest-frame optical imaging, we can compare where star formation occurred more recently in the galaxy versus where it occurred in the past.

High spatial resolution {\ha} emission-line maps can be obtained through three approaches at both low and high redshift: narrow-band imaging covering the wavelength of {\ha}, integral field unit (IFU) spectroscopy, and space-based slitless (or grism) spectroscopy. Detailed studies of the outside-in shut-down of star formation (or quenching) in local cluster galaxies were made possible with the narrow-band imaging technique (see e.g. A Virgo Environmental Survey Tracing Ionised Gas Emission (VESTIGE), \citealt{Fossati2018,Boselli2020,Boselli2021}). In the low-redshift Universe, large IFU surveys such as Mapping Nearby Galaxies at Apache Point Observatory (MaNGA; \citealt{Bundy2015}) have enabled measurements of the {\ha} spatial distribution in $\gtrsim10,000$ nearby galaxies (e.g. \citealt{Belfiore2017}). At higher redshift ($z\sim1.7$) but with lower spatial resolution, the IFU survey KMOS$^{\mathrm{3D}}$ \citep{Wisnioski2015,Wisnioski2019} confirmed the inside-out growth picture by measuring larger {\ha} sizes than the rest-frame optical \citep{Wilman2020} in star-forming galaxies that were pre-selected from the 3D-HST survey \citep{VanDokkum2011,Brammer2012,Momcheva2016}. Space-based slitless spectroscopy has added advantages over other techniques. Firstly, it provides high spatial resolution two-dimensional spectra for all the sources in the field of view, requiring no pre-selection and providing an unbiased sample of galaxies with {\ha} emission. Secondly, it can provide {\ha} emission-line maps over a larger wavelength (and ultimately, redshift) range than those provided by narrow-band imaging. \cite{Nelson2012} and \cite{ Nelson2016} were the first to use this approach, providing the first evidence of inside-out growth of star formation at high redshift ($z\sim1$) using {\ha} maps of 3200 galaxies across the star formation main sequence (SFMS; \citealt{Popesso2023} and references therein). Later, the same technique was used to provide the first direct evidence of rapid outside-in environmental quenching in galaxy clusters \citep{Matharu2021} at the same redshift. Expanding upon these works, \cite{Matharu2022} made and combined measurements from the CLEAR \citep{Estrada-Carpenter2018,Simons2023}, 3D-HST, and KMOS$^{\mathrm{3D}}$ surveys to provide the first comparison of spatially resolved star formation in star-forming galaxies over multiple epochs between $0.5\lesssim z \lesssim 1.7$ traced by {\ha} emission-line maps. These authors found a $(19\pm2)\%$ higher suppression of on-going star formation in the central kiloparsec of $z\sim0.5$ star-forming galaxies compared to $z\sim1$ star-forming galaxies at fixed stellar mass, citing the increased significance of inside-out quenching at later times for their result.

Prior to the launch of the {\it JWST}, spatially resolved studies of on-going star formation using {\ha} emission-line maps from space-based slitless spectroscopy were limited to $0.22 < z < 1.5$ due to the wavelength ranges of the blue (G102) and red (G141) grisms on the {\it HST} WFC3. With the first space-based IFUs \citep{Boker2023,Gardner2023,Hutchison2023,Rigby2023,Rigby2023b,Wright2023}, {\it JWST} has already enabled the first spatially resolved measurements of star formation, active galactic nucleus (AGN) activity, dust attenuation, and gas-phase metallicity in small samples of galaxies at $z>3$ using multiple emission lines \citep{Arribas2023,Birkin2023,DEugenio2023,Perna2023, Parlanti2023, Pino2023, Ubler2023,Ubler2023b, Jones2024, Loiacono2024, Saxena2024,Wang2024}. The slitless spectroscopic capabilities on the {\it JWST} {\it Near-infrared Imager and Slitless Spectrograph} ({NIRISS}, \citealt{Willott2022}) and the {\it Near Infrared Camera} ({NIRCam}, \citealt{Rieke2003,Rieke2005}) have now transformed the landscape of spatially resolved studies with slitless spectroscopy. {\it NIRISS} has already provided the first spatially resolved dust attenuation profiles of galaxies at $1.0<z<2.4$ using {\ha} and H$\upbeta$ emission-line maps \citep{Matharu2023}. The first high spectral and spatial resolution grism in space on NIRCam has led to the first spatially resolved rest-optical kinematic measurements of a rapidly rotating galaxy at $z=5.3$ \citep{Nelson2023}.

In this paper, we extend the redshift baseline of spatially resolved studies of star formation using {\ha} emission-line maps out to $4.8<z<6.5$ with {\it JWST} NIRCam F444W imaging and slitless spectroscopy from the First Reionization Epoch Spectroscopically Complete Observations (FRESCO; \citealt{Oesch2023}). Our sample and the data processing is described in Section~\ref{sec:sample}. A detailed description of our unique measurement strategy is provided in Section~\ref{sec:strategy}, and our results are presented in Section~\ref{sec:results}. The physical interpretation of our results and their comparison to works at similar and lower redshift are discussed in Section~\ref{sec:discussion}. A summary of our findings is given in Section~\ref{sec:summary}.

All magnitudes quoted are in the AB system, logarithms are in base 10, and we assume a $\Lambda$CDM cosmology with $\Omega_{m}=0.307$, $\Omega_{\Lambda}=0.693$, and $H_{0}=67.7$~kms$^{-1}$~Mpc$^{-1}$ \citep{Planck2015}.


\section{Sample}
\label{sec:sample}
\subsection{Data and data processing}
\label{sec:data}
Our data come from FRESCO, which is a 53.8-hour JWST Cycle 1 medium program covering 60.4 arcmin$^{2}$ in each of the GOODS-S and GOODS-N CANDELS fields with $\sim2$-hour deep F444W NIRCam imaging and slitless spectroscopy \citep{Oesch2023}. FRESCO provides high spectral resolution ($R\sim1600$) slitless spectroscopy from 4 - 5$\mu\mathrm{m}$, tracing the \ha~emission line at $4.8<z<6.6$. 

The Grism Redshift and Line Analysis Software (\texttt{grizli}; \citealt{Grizli2022}) was used to process the imaging and slitless spectroscopy together. In summary, \texttt{grizli} starts by downloading the raw data from the Mikulski Archive for Space Telescopes (MAST), pre-processing it for sky subtraction, flat-fielding, cosmic rays, alignment, and astrometric corrections \citep{Gonzaga2012,Brammer2015,Brammer2016}. A basis set of template flexible stellar population synthesis models (\texttt{FSPS}; \citealt{Conroy2009,Conroy2010}) is projected to the pixel grid of the 2D grism exposures using the spatial morphology from the F444W image. The 2D template spectra are then fit to the observed spectra with non-negative least squares. A line complex template is used to break redshift degeneracies. The final grism redshift is taken to be where the $\chi^{2}$ is minimised across the grid of trial redshifts input by the user. 

\subsection{Emission-line maps}
\label{sec:maps}
Grism spectra are median-filtered to remove the continuum. A running median filter including a 12-pixel central gap is used to ensure emission lines are not subtracted. The filtering is then run again after masking pixels with significant line flux. For more details on this process, we refer to \cite{Kashino2023} and \cite{Nelson2023}. Continuum-subtracted narrow-band maps can then be created at any desired output wavelength. By deliberately selecting the wavelength of the detected emission line from the grism redshift determination process, the user can create an emission-line map. Full details on \texttt{grizli} and its standard data products (including emission-line maps) can be found in \cite{Simons2020a}, \cite{Matharu2021}, and \cite{Noirot2022}.

\subsection{Sample selection}
\label{sec:sample_selection}
For FRESCO, a selection of high- and low-redshift sources were first used to stress test the capabilities of \texttt{grizli} in grism redshift fitting and emission-line map generation. We drew our sample from the high-redshift selection, which are all the Lyman-break galaxies in \cite{Bouwens2015} with photometric and spectroscopic redshifts beyond $z=4.8$ in the FRESCO fields of view. This amounted to 2064 galaxies in GOODS-S and 1963 in GOODS-N. The \texttt{grizli}-processed spectra and emission-line maps were then quality-checked by eye for spurious emission-line detections and contamination. Those with a {\ha} detection that have a integrated signal-to-noise ratio greater than 5 were then chosen, amounting to 151 galaxies in GOODS-S and 374 in GOODS-N. The \ha~emission-line maps of these galaxies were then quality-checked by eye for 1) contamination, 2) incomplete maps, 3) high levels of noise, and 4) significantly off-centre maps. This process reduced the sample to 119 galaxies in GOODS-S and 335 in GOODS-N.

\subsection{Stellar masses and star formation rates}
\label{sec:masses_sfms}
Stellar masses were derived from simultaneously fitting the photometry and emission-line fluxes using the \texttt{Prospector} spectral energy distribution (SED) fitting code \citep{Leja2017,Leja2019,Johnson2021}, and the parameters were set as described in Section 4.1 of \cite{Naidu2022}. 

   \begin{figure}
   \centering
   \includegraphics[width=\columnwidth]{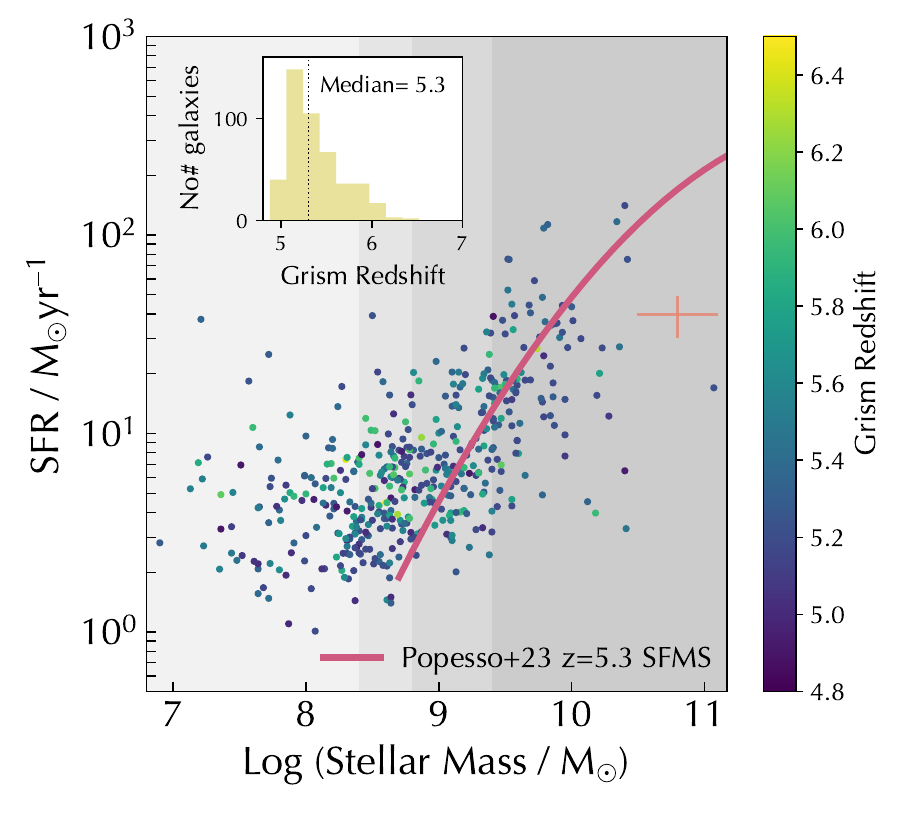}
      \caption{Star formation main sequence of our sample. The shaded grey regions delineate our stellar mass bins for the stacking. SFRs and the \cite{Popesso2023} SFMS include dust corrections (see Section~\ref{sec:masses_sfms}). The inset plot shows the grism redshift distribution of our sample. The typical measurement error is shown as the orange cross.}
              
         \label{SFMS}
   \end{figure}

The star formation rates (SFRs) were calculated using the \texttt{grizli} extracted \ha~fluxes from the grism spectra and dust-corrected using ultraviolet (UV) slopes. The UV continuum slope ($\beta$) was calculated from the \texttt{Prospector} SEDs by fitting a linear function to $\log(f_{\lambda})-\log(\lambda)$ in the UV windows defined by \cite{Calzetti1994}. The UV slopes were converted into stellar reddening ($E(B-V)$) using the conversions provided in Equation 9 of \cite{Shivaei2020} for low-metallicity galaxies. We assumed a 2.6 times higher nebular reddening compared to stellar reddening, based on the average $\frac{E(B-V)_{\rm nebular}}{E(B-V)_{\rm stellar}} = 2.6\pm0.2$ of the low-metallicity sample in \cite{Shivaei2020}. A Milky Way dust curve \citep{Cardelli1989} was used to calculate the dust attenuation of {\ha}, $A_{\rm H\upalpha}$. The \cite{Kennicutt2012} calibration was then applied to the dust-corrected \ha~luminosities to derive the SFRs. Figure~\ref{SFMS} shows the SFMS of our sample with the \cite{Popesso2023} SFMS at the median redshift of our sample overplotted in pink. Within the stellar mass range for which the \cite{Popesso2023} SFMS is valid, our sample follows the main sequence well, suggesting that we predominantly studied typical star-forming galaxies at these redshifts. At Log$(M_{*}/\mathrm{M}_{\odot})\leqslant9$, we predominantly studied galaxies with high specific star formation rates (sSFRs).


\section{Measurement strategy}
\label{sec:strategy}

   \begin{figure*}
   \centering
   \includegraphics[width=\hsize]{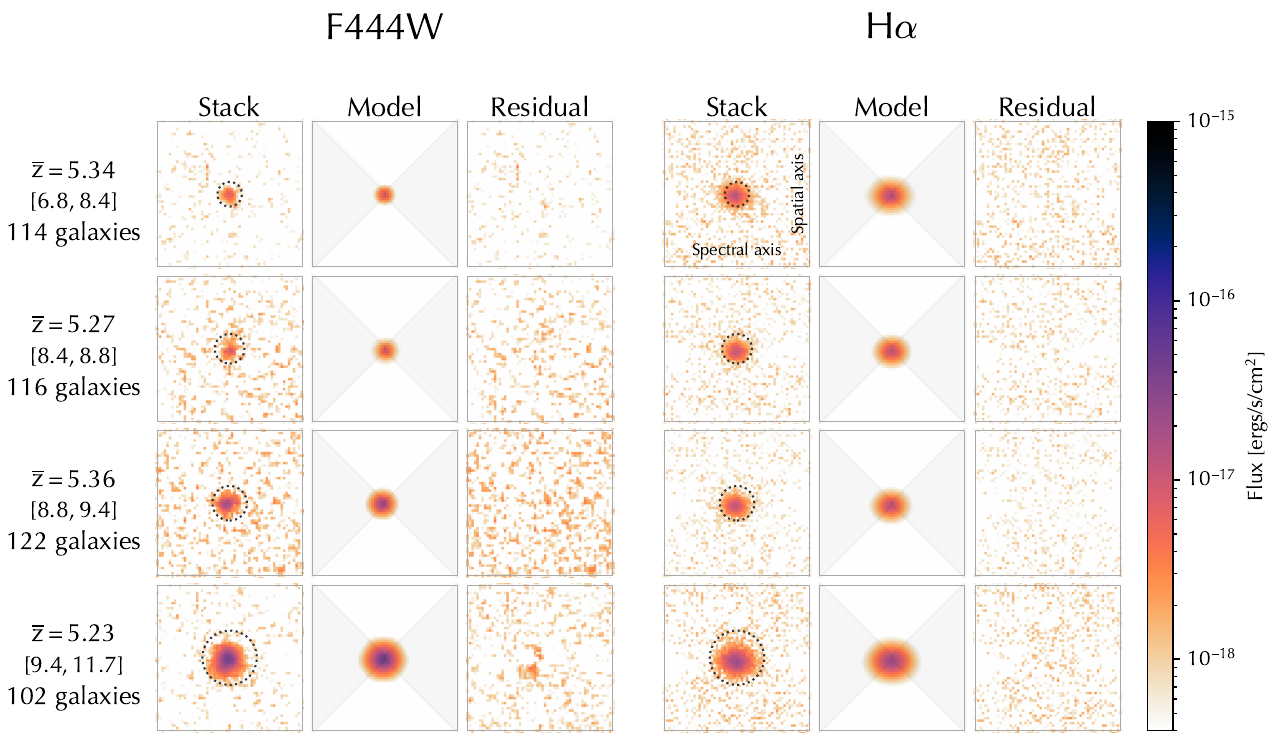}
      \caption{F444W and H$\alpha$ stacks with their associated \texttt{GALFIT} fits. Models are PSF-convolved single-component S\'ersic profiles (see Section~\ref{sec:morphology} for more details). Residuals after the model was subtracted from the data are shown in the last column for both F444W and {\ha}. Each thumbnail is $60 \times 60$ pixels, where 1 pixel = $0.05^{\prime\prime}$. $\overline{z}$ is the median grism redshift of each stack, below which the Log(M$_*/\mathrm{M}_\odot)$ range of each stack is shown in square brackets. The colour map is logarithmic, with H$\alpha$ stacks and fits multiplied by  100 for visibility. The shaded grey regions show pixels that were ignored by the hourglass mask (see Section~\ref{sec:strategy}), and the dotted black circles show the region within which surface brightness profiles were measured (see Section~\ref{sec:results}).
              }
         \label{FigStacks}
   \end{figure*}

The high spectral resolution of the NIRCam grism means that the spectral axis contains velocity information. $R\sim1600$ corresponds to a velocity dispersion of $\sigma\sim80\mathrm{km~s}^{-1}$ \citep{Nelson2023}. Velocity gradients greater than this velocity dispersion lead to a morphological distortion of emission-line maps along the spectral axis. The maximum spatial information in an emission-line map from NIRCam slitless spectroscopy is therefore obtained along the cross-dispersion axis, hereafter, the spatial axis. In this section, we describe our measurement strategy, which was designed to maximally exploit the spatial axis for spatially resolved measurements.

\subsection{Stacking}
\label{sec:stacking}
We rotated all grism spectra such that the horizontal axis was the spectral axis before we performed the redshift-fitting and extracted emission-line maps with \texttt{grizli}, as described in Section~\ref{sec:maps}. F444W thumbnails of each galaxy at the same orientation were also created as part of this \texttt{grizli} extraction process. From here on, our stacking method followed that of \cite{Matharu2022}, which was conducted on deep F105W and \ha~maps from {\it HST} WFC3 Slitless Spectroscopy. Neighbouring sources were masked in both the F444W thumbnails and \ha~emission-line maps using the \texttt{grizli}-generated segmentation maps for each F444W thumbnail. \texttt{Grizli} generated inverse variance maps for both the F444W thumbnail and the \ha~emission-line map, which we used to weight each pixel. An additional weighting by the total F444W flux was added to ensure that no single bright galaxy dominated the final stack \citep{Nelson2016}. The galaxies were separated into four bins of stellar mass: $6.8\leqslant$Log($M_{*}$/M$_{\odot}$)$<8.4$, $8.4\leqslant$Log($M_{*}$/M$_{\odot}$)$<8.8$, $8.8\leqslant$Log($M_{*}$/M$_{\odot}$)$<9.4$, and $9.4\leqslant$Log($M_{*}$/M$_{\odot}$)$<11.1$. These bins were chosen to allow for a similar number of galaxies in each stack, and they are shown in relation to our sample as the shaded grey regions in Figure~\ref{SFMS}. Within each of these bins, the F444W thumbnails and \ha~maps of each galaxy are summed and exposure-corrected using the sum of their weight maps. The variance maps for each stack are defined as $\sigma_{ij}^2 = 1/\sum{w_{ij}}$, where $w_{ij}$ is the weight map for each galaxy in the stack. The dimensions of each stack are $200\times200$ pixels with a pixel scale of $0.05^{\prime\prime}$. The zoomed-in regions of each F444W and \ha~stack are shown in Figure~\ref{FigStacks}.

\subsection{Morphology measurements}
\label{sec:morphology}
The goal of our study was to make the first spatially resolved measurements of on-going star formation compared to the stellar continuum at the highest redshifts possible with JWST Slitless Spectroscopy and place them within the context of analogous lower-redshift measurements. We therefore used the same size determination process as described in \cite{Matharu2022}, who performed the same study with {\it HST} WFC3 Slitless Spectroscopy at $z\sim0.5$ for star-forming galaxies as part of the CANDELS Lyman-Alpha Emission at Reionisation (CLEAR) survey \citep{Estrada-Carpenter2018,Simons2023}. The only modifications we made to the process were that we accounted for the loss of spatial information along the spectral axis.

We used the two-GALFIT-run approach \citep{Matharu2018}, which uses \texttt{GALFIT} \citep{Peng2002,PengGALFIT2010} to fit 2D single-component S\'ersic profiles to the stacks in two iterations. In the absence of kinematic distortions along the spectral axis, the stacking of a sufficiently large sample of galaxies leads to a stacked image with a round morphology, namely with an axis ratio equal to one. The addition of kinematic distortions leads to stacks with distinctly elongated morphologies along the spectral axis. To account for this morphological distortion, we fixed the position angle to $90$ degrees from the beginning of the fitting process. In the first iteration of the two-GALFIT-run approach, $x$, $y$ coordinates, magnitude, half-light radius, S\'ersic index, and axis ratio were kept free. The second iteration fixed the values obtained for the $x$, $y$ coordinates and the axis ratio in the first run and then re-ran \texttt{GALFIT}. Fixing the position angle at 90 degrees forces the semi-major axis to be along the spectral axis, thereby forcing \texttt{GALFIT} to measure the half-light radius along this axis. It is then straightforward to use the \texttt{GALFIT}-measured axis ratio to calculate the half-light radius along the semi-minor axis, which lies along the spatial axis.

As part of the two-GALFIT-run approach, we used a point-spread function (PSF), a sigma image, and a bad-pixel mask. The PSF accounts for the resolution limit of JWST NIRCam. \cite{Weibel2024} identified isolated stars in the FRESCO fields of view that they used as part of their PSF construction. We selected a star in GOODS-S and GOODS-N from the list of isolated stars in this work that had been through the same \texttt{grizli} extraction process as our galaxies. This ensured that they were subjected to the same data reduction process and had the same pixel scale and orientation as our galaxies. We used the same method for stacking PSFs as in Section 4.2.1 of \cite{Matharu2022}. Neighbouring sources in the PSF F444W \texttt{grizli}-generated thumbnail were masked. A mask was then created for each PSF indicating which pixels contain nonzero finite values. Each galaxy in the stack from the GOODS-S or GOODS-N field was assigned the PSF that we selected from that respective field. We then summed the PSFs for each stack and divided by the sum of their masks, effectively generating a PSF for that stack. The sigma image provides an estimate of the error per pixel for each stack and was calculated using the \texttt{grizli}-generated weight maps as described in Section~\ref{sec:stacking}. The bad-pixel mask was used for the \ha~stacks and ensured that the isolated regions of negative pixels either side of the emission-line map along the spectral axis that were created due to the median-filtering process to remove the continuum (see Section~\ref{sec:data}) were ignored by \texttt{GALFIT}. The resulting fits are shown in Figure~\ref{FigStacks}. These fits enabled us to calculate half-light radius, $R_{\mathrm{eff}}$, measurements along the spatial axis and central surface densities within one kiloparsec, $\Sigma_{1\mathrm{kpc}}$. $\Sigma_{1\mathrm{kpc}}$ is defined as

\begin{equation}
    \Sigma_{1\mathrm{kpc}} = \frac{M_{*}\gamma(2n, b_{n}R^{-1/n}_{\mathrm{eff}})}{\pi},
\end{equation}

where $M_{*}$ is the stellar mass, $n$ is the S\'ersic index, and $b_{n}$ satisfies the inverse to the lower incomplete gamma function, $\gamma(2n, b_{n})$. The regularised lower incomplete gamma function $\gamma(2n, 0.5)$ is defined as

\begin{equation}
\label{eq:lower_inc_gamma}
    \gamma(2n, 0.5) = \frac{1}{\Gamma(2n)}\int_{0}^{0.5} t^{2n-1} e^{-t} dt,
\end{equation}

where $\Gamma$ is the Gamma function. $\gamma(2n, b_{n}R^{-1/n}_{\mathrm{eff}})$ therefore takes the same form as equation~\ref{eq:lower_inc_gamma}. For more details on the $\Sigma_{1\mathrm{kpc}}$ parameter, we refer to \cite{Cheung2012a}, \cite{Barro2017a} and Section 5.2 of \cite{Matharu2022}.

\subsection{Surface brightness profiles}
\label{sec:sbprofiles}

Surface brightness profiles provide more information on the spatial distributions of \ha~and the continuum with radius than a single parametric measurement. They also provide a check on the quality of the two-dimensional models generated by \texttt{GALFIT}. \texttt{MAGPIE}\footnote{https://github.com/knaidoo29/magpie/} was used to measure the surface brightness profiles of the \ha~and continuum stacks, their PSFs, and their \texttt{GALFIT} models. The sigma images for each \ha~and continuum stack (see Section~\ref{sec:morphology}) were used in this process to calculate the errors on the surface brightness profiles. We used two different masks designed to maximise the use of the spatial axis during this measurement process. The first mask masked all pixels except for the central vertical strip along the spatial axis. These pixels are least affected by the morphological distortion along the spectral axis due to the high spectral resolution. The second mask is an hourglass-shaped mask that was designed to maximise the signal-to-noise ratio at large galactocentric radii along the spatial axis. This mask is shown as the shaded grey regions in the models in Figure~\ref{FigStacks}. It masked all pixels except for the central three pixels along the spatial axis and those within a 40 degree angle centered on the pixel above and below the central pixel.


\section{Results}
\label{sec:results}

   \begin{figure*}
   \centering
   \includegraphics[width=\textwidth]{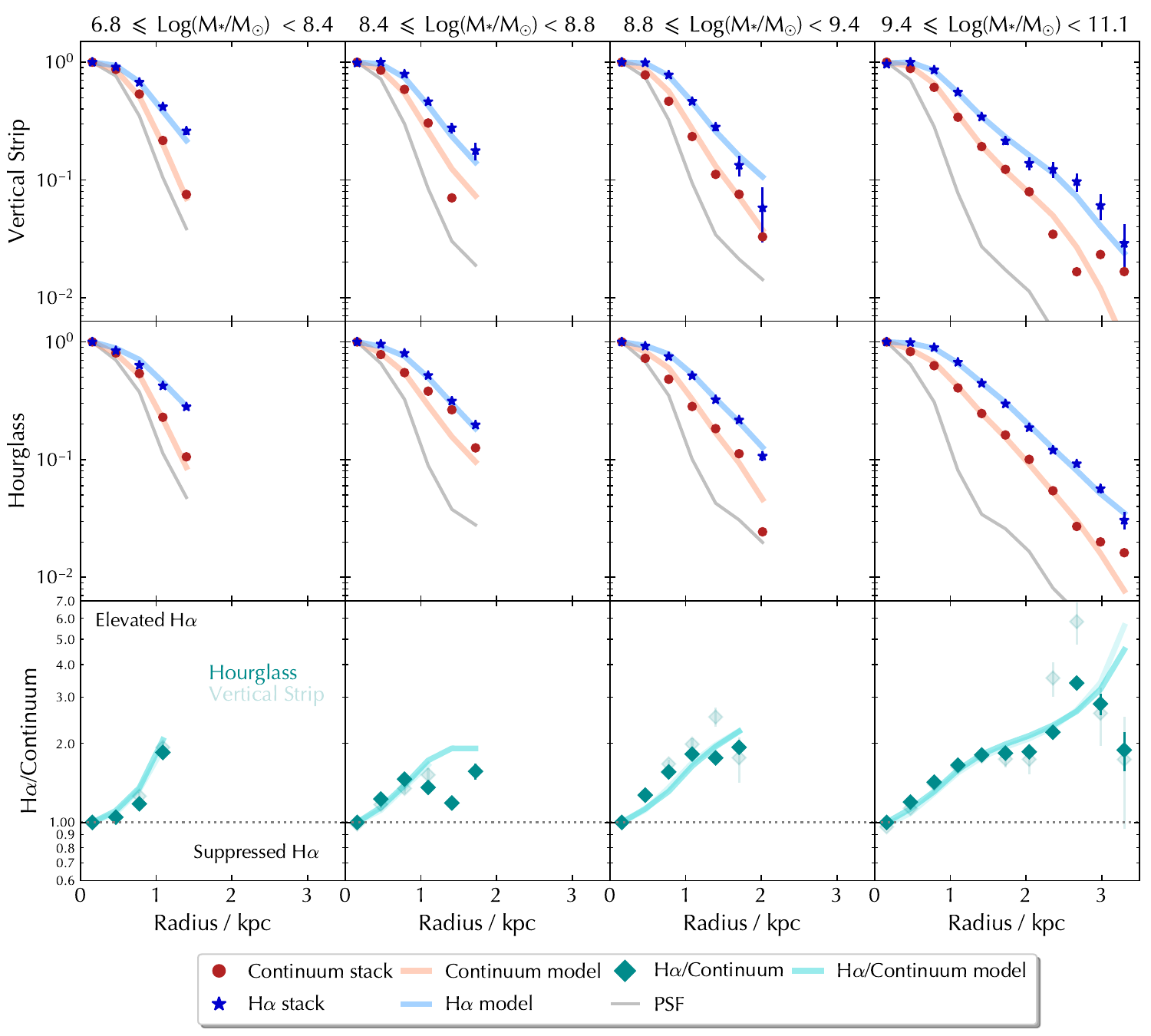}
      \caption{Peak-normalised surface brightness profiles of the FRESCO stellar continuum and H$\alpha$ stacks along with their PSF-convolved best-fit \texttt{GALFIT} models and PSFs accounting for pixels only within the vertical strip along the centre of the thumbnail (top row) or within the hourglass mask (middle row). Bottom row: Peak-normalised {\ha} equivalent width ([{\ha}/C]) profiles for both the stacks and PSF-convolved best-fit \texttt{GALFIT} models. [{\ha}/C]$\geqslant 1$ and increases with increasing galactocentric radius, suggesting the inside-out growth via star formation of star-forming galaxies as early as $z\sim5.3$.} 
              
         \label{fig:profiles}
   \end{figure*}

The peak-normalised surface brightness profiles of the stacks and their \texttt{GALFIT} models are shown in the first two rows of Figure~\ref{fig:profiles}. The first and second rows show the profiles when the vertical strip and hourglass masks were applied, respectively (see Section~\ref{sec:sbprofiles}). The PSF profiles represent the resolution limit of the telescope and are also shown for both cases in grey. The \ha~surface brightness profiles are more extended in all cases, and this is seen more explicitly in the bottom row of Figure~\ref{fig:profiles} where we show the peak-normalised {\ha} equivalent width, [{\ha}/C] profiles. These are the quotient of the {\ha} and continuum peak-normalised surface brightness profiles. [{\ha}/C]~$\geqslant1$ at all radii, reaching a maximum of $3.4\pm0.2$ for the most massive galaxies at a galactocentric radius of 2.67~kpc for the hourglass mask. 

At increasingly larger galactocentric radii for galaxies with small angular sizes such as those in our sample, the hourglass mask incorporates more pixels within which the flux is more susceptible to the morphological distortion along the spectral axis (see Section~\ref{sec:strategy}). It is therefore reassuring that at large radii, resolved measurements from using the vertical strip and hourglass masks usually agree within $1-2\sigma$ of each other. Hence, the velocity gradients at large radii in these galaxies are likely of the same order or below the velocity dispersion probed by the spectral resolution limit of the NIRCam grism (see Section~\ref{sec:strategy}). The hourglass mask is therefore an appropriate tool for increasing the signal-to-noise ratio of our measurements at large galactocentric radii.

\subsection{Evolution of spatially resolved {\ha}~versus continuum emission}
\label{sec:zevolution}

\begin{figure*}
   \sidecaption
   \includegraphics[width=12cm]{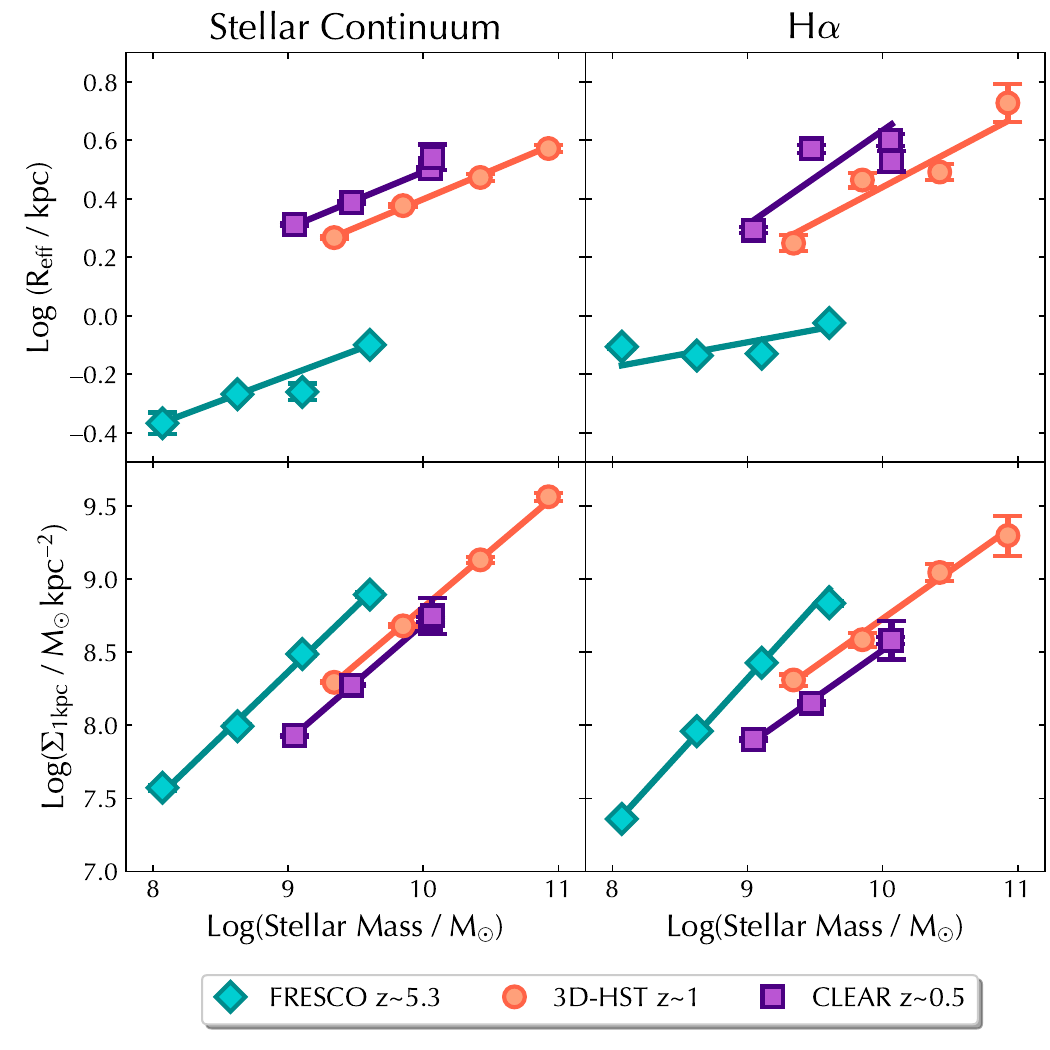}
      \caption{$R_{\mathrm{eff}}$ and $\Sigma_{1\mathrm{kpc}}$ vs. stellar mass for the stellar continuum (first column) and \ha~(second column) spatial distributions of star-forming galaxies at $z\sim5.3$ (FRESCO, this work), $z\sim1$ (3D-HST; \citealt{Nelson2016,Matharu2022}), and at $z\sim0.5$ (CLEAR; \citealt{Matharu2022}). All measurements are from a stacking analysis with {\it HST} WFC3 and {\it JWST} NIRCam slitless spectroscopy. The best-fit lines are log-linear fits, the parameters for which are given in Table~\ref{tab:lines_sep}. The stellar continuum half-light radius of star-forming galaxies at $z\sim5.3$ is 2.6 times smaller and the central kiloparsec surface density 2.5 times higher than their $z\sim1$ counterparts at a fixed stellar mass of Log(M$_*/\mathrm{M}_\odot)=9.5$. For \ha, these values are 2.3 and 2.7, respectively.
              }
         \label{FigReffContSep}
\end{figure*}

\begin{table*}  
  \centering
    \caption{Least-squares linear fits to the Log$(\mathrm{R}_{\mathrm{eff}})$ and Log$(\Sigma_{1\mathrm{kpc}})$ stellar mass relations for the stellar continuum and \ha~shown in Figure~\ref{FigReffContSep}.}
\label{tab:lines_sep}
  \begin{tabular}{|c|c|c| c c c c|}
        \cline{1-2}
      \cline{4-7}
    \multirow{2}{*}{Dataset} & Median & & \multicolumn{2}{c|}{Log($R_{\mathrm{eff}}$)} & \multicolumn{2}{c|}{Log($\Sigma_{1\mathrm{kpc}}$)} \\ 
    \cline{4-7}
    & Log$(M_{*}/\mathrm{M}_{\odot})$ & & Gradient & Intercept & Gradient & Intercept \\ \hline

    CLEAR & \multirow{2}{*}{9.47} & Cont & $0.191667\pm0.000005$ & $-1.425\pm0.004$   & $0.19400\pm0.00007$ & $-1.542\pm0.007$\\
    
    $0.22 \lesssim z \lesssim 0.75$ & & \ha & $0.32\pm0.02$ & $-2.6\pm1.5$ & $0.244\pm0.005$ & $-2.0\pm0.5$ \\ \hline
    3D-HST & \multirow{2}{*}{9.62} &Cont & $0.7966\pm0.0001$ & $0.726\pm0.009$ & $0.7828\pm0.0002$ & $0.98\pm0.02$\\
    $0.7<z<1.5$ & & \ha & $0.651\pm0.001$ & $2.0\pm0.1$  & $0.654\pm0.002$ & $2.2\pm0.2$ \\ \hline
    FRESCO & \multirow{2}{*}{8.79} & Cont & $0.1726\pm0.0001$ & $-1.76\pm0.01$ & $0.889\pm0.002$ & $0.4\pm0.2$\\
    $4.87<z<6.52$ & &\ha & $0.085\pm0.002$ & $-0.9\pm0.2$  & $1.013\pm0.001$ & $-0.79\pm0.09$\\ \hline
  
  \end{tabular}
\end{table*}









In this section, we compare our FRESCO measurements to analogous measurements made at $z\sim1$ and $z\sim0.5$ with {\it HST} WFC3 slitless spectroscopy as part of the 3D-HST \citep{Nelson2016} and CLEAR surveys \citep{Matharu2022}, respectively.

Figure~\ref{FigReffContSep} shows our $R_{\mathrm{eff}}$ and $\Sigma_{1\mathrm{kpc}}$ measurements for the stellar continuum and \ha~emission in comparison to those made at $z\sim1$ and $z\sim0.5$. The errors on these measurements come from jackknife resampling. The star-forming galaxies at $z\sim5.3$ are much smaller and more compact than their counterparts at lower redshifts when we compare their stellar continuum and \ha~distributions at fixed stellar mass. The drop in the \ha~$R_{\mathrm{eff}}$ between $z\sim1$ and $z\sim5.3$ is similar to the drop in the stellar continuum $R_{\mathrm{eff}}$ at a fixed stellar mass of Log(M$_*/\mathrm{M}_\odot)=9.5$. At the same stellar mass, the rise in $\Sigma_{1\mathrm{kpc}}$ between $z\sim1$ and $z\sim5.3$ is 1.1 times larger in \ha~than in the stellar continuum. 

\begin{figure}
   \centering
   \includegraphics[width=0.95\columnwidth]{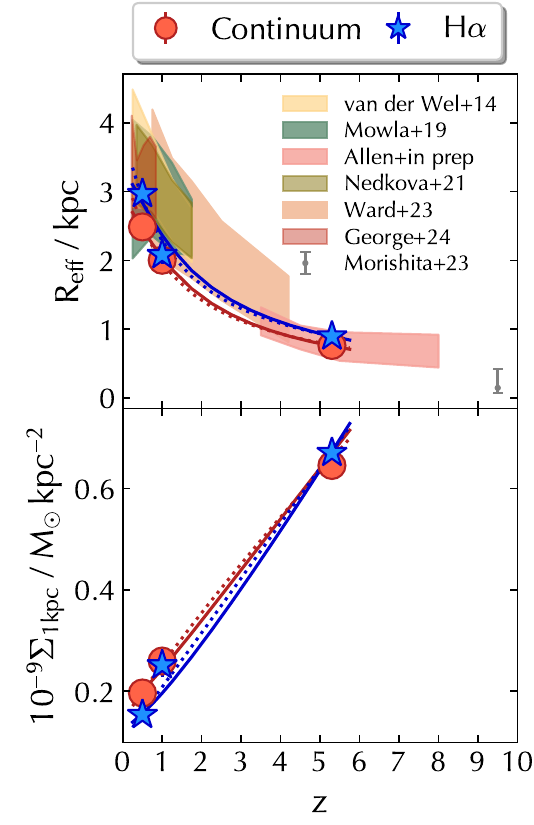}
      \caption{Morphology redshift evolution at a fixed stellar mass of Log(M$_*/\mathrm{M}_\odot)=9.5$. The solid and dotted lines are functions of the form $Ah(z)^{B}$ and $A(1+z)^{B}$, respectively, defined in Table~\ref{tab:evolution_lines}. The literature results are taken from parametrised fits to the stellar mass--size relation of star-forming galaxies. The Allen et al. (in prep) values are taken from parametrised fits to JWST NIRCam F444W imaging of star-forming galaxies in CEERS \citep{Bagley2023}, PRIMER-UDS, and PRIMER-COSMOS \citep{Dunlop2021}.}
              
         \label{fig:evolution}
   \end{figure}

The redshift evolution in the central surface density of the \ha~distribution is faster than that of the stellar continuum. This is shown more explicitly in Figure~\ref{fig:evolution}, where we use our parametrised fits shown as the solid lines in Figure~\ref{FigReffContSep} and given in Table~\ref{tab:lines_sep} to calculate $R_{\mathrm{eff}}$ and $\Sigma_{1\mathrm{kpc}}$ at a fixed stellar mass of Log(M$_*/\mathrm{M}_\odot)=9.5$ for both \ha~and the continuum. Following \cite{VanderWel2014}, we fitted functions of the form $A(1+z)^{B}$ and $Ah(z)^{B}$ to our measurements, defined and given in Table~\ref{tab:evolution_lines}. We also show continuum $R_{\mathrm{eff}}$ measurements at this fixed stellar mass from \cite{VanderWel2014}, \cite{Mowla2019}, \cite{Nedkova2021}, \cite{Morishita2023}, \cite{Ward2023}, \cite{George2024}, and Allen (et al. in prep). These were calculated using parametrised fits to the stellar mass-size relations of star-forming galaxies given in these works.

\begin{table}  
  \centering
    \caption{Parametrised fits to the redshift evolution shown in Figure~\ref{fig:evolution}.}
\label{tab:evolution_lines}

\begin{tabular}{|c|c|cc|cc| }
    \cline{3-6}
     \multicolumn{2}{c|}{} & \multicolumn{2}{c|}{$Ah(z)^{B}$} & \multicolumn{2}{c|}{$A(1+z)^{B}$} \\
    \cline{3-6}
     \multicolumn{2}{c|}{} & $A$ & $B$ & $A$ & $B$ \\
     \hline
    \multirow{2}{*}{$R_{\mathrm{eff}}(z)$} & Cont & 2.3 & -0.6 & 3.5 & -0.8 \\
    & \ha & 2.6 & -0.6 & 4.0 & -0.8  \\
    \hline
    \multirow{2}{*}{$10^{-9}\Sigma_{1\mathrm{kpc}}(z)$} & Cont & 0.2 & 0.6 & 0.1 & 0.8 \\
    & \ha & 0.2 & 0.8 & 0.1 & 1.0  \\
    \hline

\end{tabular}
\tablefoot{The Hubble parameter, $H(z)=100h(z)\mathrm{km~s}^{-1}\mathrm{Mpc}^{-1}$.}
\end{table}

\begin{figure*}
   \centering
   \includegraphics[width=17cm]{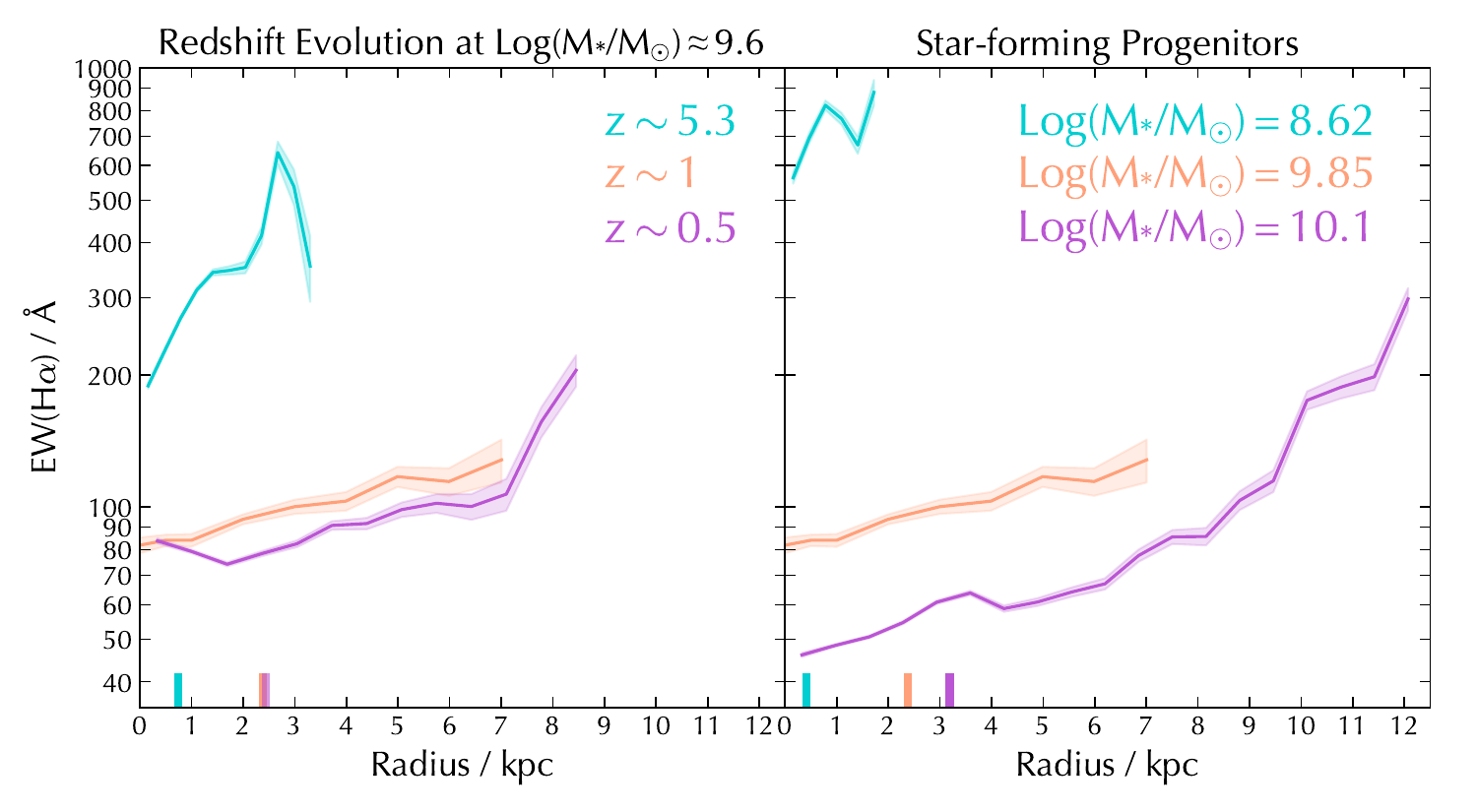}
      \caption{{\ha} equivalent width profiles in star-forming galaxies at $z\sim0.5$ \citep{Matharu2022}, $z\sim1$ \citep{Nelson2016}, and $z\sim5.3$ (this work). The small thick vertical lines show the continuum $R_{\mathrm{eff}}$ measurements for each stack. Left: At fixed stellar mass,  star-forming galaxies that form earlier grow inside-out via star formation more rapidly with galactocentric radius. Right: If a galaxy forms stars at a constant rate between $5.27<z<1$ and $1<z<0.5$, starting with a stellar mass of Log$(M_{*}/\mathrm{M}_{\odot})=8.62$ at $z=5.27$ (see Section~\ref{sec:zevolution} for details), it will rapidly grow its inner region via star formation with radius, after which there will be steady growth at $z=1$ and then rapid growth of the outer region with radius at $z=0.5$.}
              
         \label{fig:sf_evolution}
\end{figure*}

Whilst the evidence of extended \ha~relative to the continuum at all stellar masses is clear in Figure~\ref{fig:profiles}, this is not apparent when comparing the \ha~and continuum $\Sigma_{1\mathrm{kpc}}$  measurements at $z=5.3$ in Figure~\ref{fig:evolution}. We measure slightly larger $\Sigma_{1\mathrm{kpc}}$ in \ha~than in the continuum, suggesting more centrally concentrated {\ha} emission than the continuum emission within 1 kpc. These results highlight that morphology measurements from parametrised fits alone cannot provide a full picture of how the \ha~emission relative to the continuum is spatially distributed. 

Therefore, we show the evolution in {\ha} equivalent width, EW({\ha}), profiles using results from CLEAR, 3D-HST, and our work in Figure~\ref{fig:sf_evolution}. The left panel shows the redshift evolution in the EW({\ha}) profiles of star-forming galaxies at a fixed stellar mass. The EW({\ha}) profiles from these works have similar median masses (Log$(M_{*}/\mathrm{M}_{\odot})=9.6, 9.9$, and $9.5$ at $z=5.23$, $z\sim1$, and $z=0.58$, respectively). At fixed stellar mass, a star-forming galaxy that forms at earlier times will exhibit a rapid rise in EW({\ha}) with galactocentric radius. At $z\sim1$, there is a more steady rise with radius, with a rapid rise seen in the outer regions at $z\sim0.5$.

Given the median stellar masses of the four FRESCO stacks, we calculated the stellar masses of corresponding star-forming galaxies at $z=1$ and $z=0.5$ with the assumption that the SFRs of the galaxies dictated by the \cite{Popesso2023} relation at $z\approx5.3$ and $z=1$ is the SFR at which the galaxies form stars at a constant rate between $5.3\lesssim z<1$ and $1<z<0.5$, respectively. In the right panel of Figure~\ref{fig:sf_evolution}, we show the EW({\ha}) profiles whose median stellar masses most closely match those of our calculations at $z=1$ and $z=0.5$ from \cite{Nelson2016} and \cite{Matharu2022}, respectively, with the corresponding FRESCO EW({\ha}) profile at $z=5.27$. These results show that a star-forming galaxy with a stellar mass of Log$(M_{*}/\mathrm{M}_{\odot})=8.62$ will exhibit a rapid rise in EW({\ha}) with radius at $z=5.27$, after which its descendants exhibit a more suppressed rise with radius in EW({\ha}) at $z=1$ and $z=0.5$, but with an increasing rate with radius in the outer regions beyond 8~kpc at $z=0.5$.

In an absolute sense, EW({\ha}) falls more rapidly at fixed radius when star-forming progenitors are tracked than at fixed mass. At a fixed $r=1$~kpc, the drop in  EW({\ha}) between $z\sim5.3$ and $z\sim1$ is $3.0\pm0.1$ times larger for star-forming progenitors than at fixed mass, and this factor increases to $7\pm4$ for the same comparison between $z\sim1$ and $z\sim0.5$. Furthermore, the suppression of EW({\ha}) with radius is fairly uniform between $z\sim1$ and $z\sim0.5$ within $r=7$~kpc at fixed mass and when star-forming progenitors are tracked.

In the next section, we discuss the physical implications of our results.


\section{Discussion}
\label{sec:discussion}
The youngest, most massive stars of type O with lifetimes of $\sim 10$~Myr emit strong UV radiation capable of stripping hydrogen atoms of their electrons. Recombination at the boundaries of their Str\"{o}mgren spheres \citep{Stromgren1939} leads to the emission of \ha. \ha~emission therefore predominantly traces on-going star formation over the last 10~Myr. Thus,   Figure~\ref{fig:profiles} provides the first spatially resolved measurements of on-going star formation across the star-forming main sequence (Figure~\ref{SFMS}) at $z\sim5.3$.  The strength of the \ha~emission was calculated by taking its quotient with the underlying continuum, known as the \ha~equivalent width. This quantity has the added benefit of dividing out the diffuse dust attenuation that affects the \ha~and continuum emission equally (\citealt{Nelson2016}, but see Section~\ref{sec:dust} for further discussion). Therefore, the peak-normalised \ha~equivalent width profiles shown in the bottom row of Figure~\ref{fig:profiles} allowed us to measure how suppressed or elevated on-going star formation is relative to the underyling continuum that is dominated by the emission from older stars, and how this compares to the galaxy centre. We interpret \ha~emission below as on-going star formation and the continuum as the integrated star formation history. 

We find that star formation is elevated at all galactocentric radii relative to the centre for all star-forming galaxies with $6.8\leqslant$Log($M_{*}$/M$_{\odot}$)$<11.1$. Importantly, [{\ha}/C] has an overall positive trend with galactocentric radius, consistent with increasingly elevated star formation with distance from the centre of these galaxies. This suggests that star-forming galaxies just after the epoch of reionisation grow inside-out via star formation.

\subsection{Evolution of spatially resolved star formation with cosmic time}

Including the study presented in this paper, {\it HST} and {\it JWST} slitless spectroscopy has now led to direct measurements of spatially resolved star formation using \ha~emission in star-forming galaxies between $0.22<z<6.5$ \citep{Nelson2016a, Matharu2022}. This provides the first opportunity of studying the differences in how star formation proceeds in galaxies across cosmic time out to the end of reionisation using measurements made with the same technique. In this section, we discuss the interpretations that can be drawn by comparing measurements from these three studies.

\subsubsection{Bulge versus disk growth}
\label{sec:bulge_v_disk}

Figure~\ref{FigReffContSep} shows us what we expect in the evolution of galaxy morphology with redshift: High-redshift star-forming galaxies are smaller and have more concentrated light within 1 kpc than their counterparts at low redshift. At $z\sim5.3$, the stellar mass-$R_{\mathrm{eff}}$ relation is shallower in \ha~than the stellar continuum (Table~\ref{tab:lines_sep}). When \ha~is interpreted as on-going star formation and the stellar continuum as the integrated star formation history, this suggests that on-going star formation at $z\sim5.3$  progresses at a slower rate with galactocentric radius as a function of stellar mass than it did at $z<5.3$. However, the stellar mass-$\Sigma_{1\mathrm{kpc}}$ relation is steeper in \ha~than the stellar continuum. This suggests that as a function of stellar mass, the central regions of star-forming galaxies at $z\sim5.3$ form stars more rapidly than they did at $z<5.3$. At $z\sim0.5$ and $z\sim1$, this trend reverses. The stellar mass-$R_{\mathrm{eff}}$ relations in \ha~are steeper than those in the continuum, and the stellar mass-$\Sigma_{1\mathrm{kpc}}$ relations are shallower in \ha~than in the continuum (Table~\ref{tab:lines_sep}). At $z\sim5.3$, the half-light radius is always smaller than 1 kpc, and so is within the region within which $\Sigma_{1\mathrm{kpc}}$ is measured. The $R_{\mathrm{eff}}$ and $\Sigma_{1\mathrm{kpc}}$ measurements at this redshift are therefore likely both dominated by the bulge region of these galaxies. This is not true at $z\leqslant 1$. $R_{\mathrm{eff}}$ measurements at $z\sim5.3$ are more sensitive to physical processes in the central regions of star-forming galaxies than it is at $z\leqslant 1$, where it traces physical processes that occur at larger galactocentric radii. We interpret the dense central concentration as the bulges of these galaxies and the almost exponential part of their surface brightness profiles as their disks. These trends therefore tell us that star-forming galaxies at $z\sim5.3$ build their bulges rapidly. At $z\leqslant1$, star-forming galaxies no longer build their bulges rapidly via star formation and build their disks at a steady pace.

Further clarity for this picture is provided in Figure~\ref{fig:evolution}, where we show the redshift evolution in $R_{\mathrm{eff}}$ and  $\Sigma_{1\mathrm{kpc}}$ at a fixed stellar mass of Log$(M_{*}/\mathrm{M}_{\odot})=9.5$. At $z=5.3$, $\Sigma_{1\mathrm{kpc, H}\upalpha}$ is $1.04\pm0.05$ times higher than $\Sigma_{1\mathrm{kpc, C}}$. Towards lower redshift, $\Sigma_{1\mathrm{kpc, H}\upalpha}\lesssim \Sigma_{1\mathrm{kpc, C}}$. The faster decline in \ha~stellar mass surface density within 1 kpc (bottom panel of Figure~\ref{fig:evolution} and Table~\ref{tab:evolution_lines}) is therefore driven by a switch from rapid bulge growth to significant disk growth towards lower redshifts.

\subsubsection{Evolution in \ha~equivalent width profiles}
\label{sec:ew_profiles}
Whilst parametrised morphology measurements are useful for understanding the build-up of galaxies with redshift, they are sensitive to different regions of galaxies at different times (Section~\ref{sec:bulge_v_disk}). A full picture of how spatially resolved star formation proceeds as a function of galactocentric radius with redshift can be gained by comparing \ha~equivalent width profiles. We compared them at fixed stellar mass (left panel of Figure~\ref{fig:sf_evolution}) and by tracing star-forming progenitors between $5.3\lesssim z \leqslant 0.5$ (right panel of Figure~\ref{fig:sf_evolution}; more details in Section~\ref{sec:zevolution}).

All the EW({\ha}) profiles shown are positive, indicating the prevalence of inside-out growth via star formation in star-forming galaxies between $0.5\leqslant z \leqslant5.3$. However, the steep rise with radius for $z\sim5.3$ galaxies suggests that this growth occurred more rapidly as a function of radius in the early Universe. The similar slopes of the $z\sim1$ and $z\sim0.5$ EW({\ha}) profiles in each panel suggests that there is a spatially uniform suppression of star formation in star-forming galaxies between $0.5\leqslant z \leqslant1$ both at fixed mass and when star-forming progenitors are traced. At lower redshifts, significantly elevated star formation as a function of radius is seen in the outer disk of star-forming galaxies.

The redshift evolution in the EW({\ha}) profiles at fixed mass shown in the left panel of Figure~\ref{fig:sf_evolution} supports our interpretation of the redshift evolution in morphology measurements at fixed mass discussed in Section~\ref{sec:bulge_v_disk}. Star-forming galaxies just after the epoch of reionisation rapidly build their inner regions or bulges, which dominate the half-light radius (see the turquoise vertical line in the left panel of Figure~\ref{fig:sf_evolution}). This leads to compact {\ha} in the central region, driving up $\Sigma_{1\mathrm{kpc, H}\upalpha}$ relative to $\Sigma_{1\mathrm{kpc, C}}$ (bottom panel of Figure~\ref{fig:evolution}). A similar evolution is seen for star-forming progenitors (right panel of Figure~\ref{fig:sf_evolution}), the main difference being the larger drop in EW({\ha}) at fixed radius with redshift. However, it should be noted that at log$(M_{*}/\mathrm{M}_{\odot})\lesssim9$, we primarily sampled star-forming galaxies with high specific star formation rates at $z\sim5.3$ (see Figure~\ref{SFMS} and Section~\ref{sec:masses_sfms}), and so the large absolute difference in EW({\ha}) at $5.3\lesssim z \lesssim 1$ might be driven by this sample bias. 

Nevertheless, the overall picture provided by the EW({\ha}) profiles of star-forming galaxies since the epoch of reionisation is that bulges are first built rapidly with radius, after which there is a gradual inside-out cessation of star formation that flattens the EW({\ha}) profiles at $z\lesssim1$, with significantly elevated levels of inside-out growth via star formation in the outer disk at $z\sim0.5$.

\subsection{Literature comparison}

In this section, we compare our results to those in the literature, starting with a focus on results just after the epoch of reionsation (Section~\ref{sec:z5}) and continuing with results at lower redshift (Section~\ref{sec:zlow}). 

\subsubsection{Just after the epoch of reionisation}
\label{sec:z5}

Reassuringly, our F444W $R_{\mathrm{eff}}$ measurement at fixed stellar mass agrees within $1\sigma$ of the F444W measurements at the same epoch made by Allen (in prep.; top panel of Figure~\ref{fig:evolution}). Interestingly, these authors report that the rest-optical and rest-UV $R_{\mathrm{eff}}$ agree within $1-2\sigma$ of each other. Since the UV is more susceptible to dust attenuation than \ha~(\citealt{Kennicutt2012} and references therein), this result suggests that dust obscuration at these stellar masses may not be significant (see the more detailed discussion in Section~\ref{sec:dust}), and that as a result of the young stellar populations at this epoch, the rest-UV may be an appropriate tracer of the stellar mass distribution.

The {\it JWST} NIRCam grism is the first high spectral resolution grism in space (see Section~\ref{sec:strategy}), introducing the capability of making spatially resolved kinematic measurements for large samples of distant galaxies. \cite{Nelson2023} took advantage of this capability and showed that a log$(M_{*}/\mathrm{M}_{\odot})=10.4$ star-forming galaxy at $z=5.3$ has a rotating \ha~disk with $V_{rot}=240\pm50~\mathrm{kms}^{-1}$. In a low-mass (log$(M_{*}/\mathrm{M}_{\odot})=8.6\pm0.1$) galaxy at $z=8.34$, \cite{Zihao2023} measured a rotational velocity of $V_{rot}=58^{+53}_{-35}~\mathrm{kms}^{-1}$ using the [O III]$\lambda5007$ emission-line map. Both these works suggested that emission-line galaxies over a wide range in stellar mass with rotating gaseous disks exist just after the epoch of reionisation. This is despite growing evidence that the fraction of rotation-dominated star-forming galaxies falls significantly at $z\gtrsim3$ (\citealt{Turner2017,Wisnioski2019} and references therein). We discuss this trend within the context of lower-redshift studies in the next section.

\subsubsection{Dominance of disks}
\label{sec:zlow}

By extending spatially resolved studies of star formation with \ha~emission-line maps out to $z\sim5.3$, we enabled a study of the evolution in how star formation proceeds over a longer time baseline. The importance of this extended baseline is evident in Figure~\ref{fig:evolution}, which suggests a shift in predominantly bulge growth to the dominance of disk growth via star formation towards lower redshifts (see Section~\ref{sec:bulge_v_disk}), which is supported by the EW({\ha}) profiles in Figure~\ref{fig:sf_evolution}. This is consistent with findings that the fraction of disk-dominated star-forming galaxies falls significantly towards higher redshifts, both in terms of kinematics measuring rotation (see Section~\ref{sec:z5} and e.g. \citealt{ForsterSchreiber2009,ForsterSchreiber2018,Wisnioski2015,Wisnioski2019,Gnerucci2011,Kassin2012,Miller2012,Tacconi2013,Simons2017,Turner2017,Johnson2018,Ubler2019,Price2020}) and the spiral fraction \citep{Kuhn2023}.

\cite{Matharu2022} measured and compared \ha~and continuum $R_{\mathrm{eff}}$ and $\Sigma_{1\mathrm{kpc}}$ measurements from the 3D-HST and CLEAR surveys, the results of which are also shown in Figures~\ref{FigReffContSep} and \ref{fig:evolution} of this paper. At a fixed stellar mass of Log$(M_{*}/\mathrm{M}_{\odot})=9.5$, they observed significant evolution in $\frac{\Sigma_{1\mathrm{kpc,H}\upalpha}}{\Sigma
_{1\mathrm{kpc,C}}}$, where it is $(19\pm2)\%$ lower at $z\sim0.5$ than at $z\sim1$.  To explain the evolution seen between $0.5\leqslant z \leqslant1$, they referred to the natural consequences of inside-out growth: the inside-out cessation of star formation follows the inside-out growth and becomes significant at lower redshift. Figure~\ref{fig:sf_evolution} and the discussion in Section~\ref{sec:ew_profiles} confirms this picture both at fixed mass (left panel) and for progenitors on the main sequence (right panel). At $z\sim5.3$, star-forming galaxies grow their bulges rapidly, but at lower redshift, the most significant growth via star formation is only in the outer disk (Figure~\ref{fig:sf_evolution}). Our results are consistent with those of \cite{Shen2023b}, who reported that the EW({\ha}) profiles of 19 star-forming galaxies between $0.6<z<2.2$ measured using {\it JWST} NIRISS slitless spectroscopy are predominantly positive, which is in line with the inside-out growth scenario. With spatially resolved SED fitting, they further determined that the star formation histories (SFHs) of the central regions of star-forming galaxies at this epoch are consistent with having experienced at least one rapid star formation episode that led to the formation of the bulge. The disks of the galaxies in this work were found to grow with more smoothly varying SFHs.

\subsection{Effects of dust attenuation}
\label{sec:dust}

If there is increased dust attenuation towards stellar birth clouds at $z\sim5.3$, it can further suppress \ha~emission (e.g. \citealt{Calzetti1999,ForsterSchreiber2009,Yoshikawa2010,Mancini2011,Wuyts2011,Wuyts2013, Kashino2013, Kreckel2013,Price2014,Reddy2015,Bassett2017, Theios2019,Koyama2019a,Greener2020, Wilman2020,Rodriguez-Munoz2021}). If this increased nebular attenuation is centrally concentrated, it might cause the positive trends in the peak-normalised \ha~equivalent width profiles we show in Figure~\ref{fig:profiles}. To determine whether this is the case, we would need to calculate spatially resolved Balmer decrements for these galaxies, from which we can derive nebular attenuation profiles. This would require both \ha~and H$\upbeta$ emission-line maps. To obtain H$\upbeta$ emission-line maps for our sample, we require F356W NIRCam grism observations of our galaxies. These observations now exist in GOODS-N \citep{Egami2023}, but are yet to be obtained for GOODS-S. Future work will include measuring Balmer decrement radial profiles for the galaxies studied in this paper. 

Recent work on the redshift evolution of spatially resolved Balmer decrements in emission-line galaxies with {\it JWST} NIRISS slitless spectroscopy between $1.0<z<2.4$ shows no significant difference in the integrated Balmer decrements or in the shape of Balmer decrement radial profiles with increasing redshift \citep{Matharu2023}. Calculating nebular dust attenuation profiles from these assuming a \cite{Calzetti1999} dust law, \cite{Matharu2023} reported centrally concentrated nebular dust attenuation profiles for $7.6\leqslant$Log($M_{*}$/M$_{\odot}$)$<10$ emission-line galaxies, which reached a maximum attenuation towards \ha-emitting regions, $A_{\mathrm{H}\upalpha}=1.03\pm0.04$~mag within 0.4~kpc from the centre for  $9.0\leqslant$Log($M_{*}$/M$_{\odot}$)$<10.0$. Flat profiles are measured at $10\leqslant$Log($M_{*}$/M$_{\odot}$)$<11.1$, however. The results presented in \cite{Liu2023} agree with this. These authors used {\it JWST} MIRI F770W and NIRCam F200W imaging of mid-infrared selected galaxies at $1.0<z<1.7$ to decompose their dust and stellar components. They reported that $9.5<$Log($M_{*}$/M$_{\odot}$)$<10.5$ galaxies have dust cores that are $\sim1.23-1.27$ times more compact than their stellar cores, but for the most massive galaxy at Log$(M_{*}/\mathrm{M}_{\odot})=10.9$, this factor decreases to $\sim0.86-0.89$. In contrast, \cite{Shen2023a} find increasing dust obscuration fractions with stellar mass for star-forming galaxies at $0.2<z<2.5$ for the entire galaxy and when focusing only on the inner 1 kpc. The obscured fractions are lowest at Log$(M_{*}/\mathrm{M_{\odot}})\leqslant9.5$ and highest at Log$(M_{*}/\mathrm{M_{\odot}})\geqslant10.0$.
Similarly, based on {\it HST} WFC3 slitless spectroscopy, \cite{Nelson2016a} reported an increasing central concentration of nebular dust attenuation profiles with stellar mass at $z\sim1.4$. They determined $A_{\mathrm{H}\upalpha}\sim0.8$~mag for $9.0\leqslant$Log($M_{*}$/M$_{\odot}$)$<9.2$ but increased to $\sim3.5$~mag for $9.8\leqslant$Log($M_{*}$/M$_{\odot}$)$<11$ under the assumption of a \cite{Calzetti1999} dust law. 

Whilst there is no resolved Balmer decrement work at \mbox{$4.8<z<6.5$} for star-forming galaxies, dust continuum observations with the {\it Atacama Large Millimeter/submillimeter Array} (ALMA) have revealed a diversity of dust profiles at $z\sim7$ \citep{Bowler2022} and similar spatial extents for dust-obscured and unobscured star formation at $4\leqslant z \leqslant 6$ \citep{Mitsuhashi2023}. Similar spatial extents supports the probability of similarly shaped radial profiles. If the dust-obscured star formation radial profile has the same shape as the unobscured star formation radial profile, this implies that the reddening of the stellar continuum is uniform with galactocentric radius and that the nebular attenuation radial profile may be flat. Therefore, even if there is increased dust attenuation towards stellar birth clouds at $z\sim5.3$, it is unlikely to be driving the positive trends seen in our [{\ha}/C] profiles.

Collectively, these works do not provide a clear picture of how spatially resolved nebular attenuation evolves with redshift and stellar mass. Future work focused on measuring the nebular attenuation profiles of main-sequence galaxies with the same method and selected in the same way out to the epoch of reionisation will bring clarity to this problem.

\subsection{[N II] contribution to \ha}

The high spectral resolution of the NIRCam grism means that the \ha~emission line is separated from its two neighbouring [N II] emission lines by more than a resolution element\footnote{1 NIRCam pixel = $10{\AA}$}. This means that the centroids of the \ha~and [N II] emission-line maps will not overlap, but any spatially extended emission could. \cite{Faisst2018} showed that [N II]/([N II] + \ha) declines with increasing redshift for galaxies with stellar masses $8.5\leqslant$Log($M_{*}$/M$_{\odot}$)$<10$ between $0<z<2.6$, with higher-mass galaxies experiencing a modest rise between $0<z\lesssim1.5$, after which there is a steep  decline. By extrapolating their [N II]/([N II] + \ha) values linearly out to $z=5.3$ for Log$(M_{*}/\mathrm{M}_{\odot})=8.5$ and Log$(M_{*}/\mathrm{M}_{\odot})=11.1$\footnote{At $z\geqslant2$ where it begins to decline.}, we find that [N II]/([N II] + \ha) is consistent with zero for galaxies with Log$(M_{*}/\mathrm{M}_{\odot})=11.1$ and 0.02 for galaxies with Log$(M_{*}/\mathrm{M}_{\odot})=8.5$. This fraction is so small that its effects are likely within our uncertainties. Furthermore, we excluded the region of the \ha~emission-line maps where the [N II] centroids would reside using the hourglass mask, and we thereby ensured that the most severe [N II] contamination does not affect our final measurements.


\section{Summary}
\label{sec:summary}
Using the largest sample of star-forming galaxies just after the epoch of reionisation for which there exists {\it JWST} NIRCam slitless spectroscopy, we have made the first measurements on the spatial distribution of star formation traced by \ha~emission (Section~\ref{sec:results}).

To allow for direct comparisons with lower redshift works using {\it HST} WFC3 slitless spectroscopy, we processed and stacked our data using the same method whilst taking the high spectral resolution of the NIRCam grism into account (Section~\ref{sec:strategy}). 

Our main conclusions are listed below.

   \begin{enumerate}

      \item Star-forming galaxies across the main sequence at $4.8<z<6.5$ with stellar masses $6.8\leqslant \mathrm{log}(M_{*}/\mathrm{M_{\odot}})<11.1$ have positive peak-normalised EW({\ha}) profiles. This provides direct evidence for the inside-out growth of star-forming galaxies just after the epoch of reionisation.

      \item Parametrised morphological measurements at a fixed stellar mass of log$(M_{*}/\mathrm{M}_{\odot})=9.5$ reveal that whilst the stellar mass surface density within 1~kpc of {\ha}, $\Sigma_{1\mathrm{kpc, H}\upalpha}$, is $1.04\pm0.05$ times more concentrated than that of the continuum, $\Sigma_{1\mathrm{kpc, C}}$, the {\ha} half-light radius, $R_{\mathrm{eff, H}\upalpha}$ is $1.18\pm0.03$ times more extended than that of the continuum half-light radius, $R_{\mathrm{eff, C}}$, but both are smaller than 1 kpc. These results suggest the rapid build-up of compact bulges via star formation just after the epoch of reionisation (Figure~\ref{fig:evolution} and Section~\ref{sec:bulge_v_disk}).

      \item The pace of redshift evolution at a fixed stellar mass of Log$(M_{*}/\mathrm{M}_{\odot})=9.5$ for $R_{\mathrm{eff}}(z)$ is the same for {\ha} and the continuum. For $\Sigma_{1\mathrm{kpc}}(z)$, however, the redshift evolution is faster for {\ha}, with $\Sigma_{1\mathrm{kpc,~ H}\upalpha/\mathrm{C}}=h(z)^{1.3}$ (Figure~\ref{fig:evolution} and Table~\ref{tab:evolution_lines}).
      
      \item Towards lower redshifts, $\Sigma_{1\mathrm{kpc, H}\upalpha} \geqslant \Sigma_{1\mathrm{kpc, C}}$ becomes $\Sigma_{1\mathrm{kpc, H}\upalpha}\lesssim \Sigma_{1\mathrm{kpc, C}}$. This is consistent with the inside-out growth of the disk via star formation dominating the inside-out growth of the bulge at later times.
      
      \item The dominance of disk growth over bulge growth at lower redshift is further supported by the evolution in EW({\ha}) profiles at fixed stellar mass with redshift and when star-forming progenitors are traced between $5.3\lesssim z \lesssim0.5$. Star-forming galaxies at $z\sim5.3$ have rapidly increasing EW({\ha}) with radius within their half-light radius, whilst a significantly increasing EW({\ha}) with radius at $z\sim0.5$ is only seen in the outer disk.

   \end{enumerate}

Our work demonstrated that spatially resolved studies of emission-line galaxies can be conducted with NIRCam slitless spectroscopy out to the epoch of reionisation. Future work will focus on exploiting this technique to study a variety of spatially resolved physical properties of high-redshift galaxies beyond just star formation.

\begin{acknowledgements}
      JM is grateful to the Cosmic Dawn Center for the DAWN Fellowship. JM thanks Adam Muzzin, Viola Gelli and Anne Hutter for useful discussions that led to improvements in the analysis presented in this paper. 
      This work is based on observations made with the NASA/ESA/CSA James Webb Space Telescope. The raw data were obtained from the Mikulski Archive for Space Telescopes at the Space Telescope Science Institute, which is operated by the Association of Universities for Research in Astronomy, Inc., under NASA contract NAS 5-03127 for \textit{JWST}. These observations are associated with JWST Cycle 1 GO program \#1895. Support for program JWST-GO-1895 was provided by NASA through a grant from the Space Telescope Science Institute, which is operated by the Associations of Universities for Research in Astronomy, Incorporated, under NASA contract NAS5-26555. 
      The Cosmic Dawn Center (DAWN) is funded by the Danish National Research Foundation under grant DNRF140.
      This work has received funding from the Swiss State Secretariat for Education, Research and Innovation (SERI) under contract number MB22.00072, as well as from the Swiss National Science Foundation (SNSF) through project grant 200020\_207349.
      RPN thanks the NASA Hubble Fellowshp Program for the Hubble Fellowship. DM acknowledges funding from JWST-GO-01895.013, provided through a grant from the STScI under NASA contract NAS5-03127.
\end{acknowledgements}


\bibliography{library_grizli2}{}

\begin{thebibliography}{136}
\expandafter\ifx\csname natexlab\endcsname\relax\def\natexlab#1{#1}\fi

\bibitem[{Abramson {et~al.}(2016)Abramson, Kenney, Crowl, \& Tal}]{Abramson2016b}
Abramson, A., Kenney, J., Crowl, H., \& Tal, T. 2016, AJ, 152, 32

\bibitem[{Abramson {et~al.}(2011)Abramson, Kenney, Crowl, Chung, van Gorkom, Vollmer, \& Schiminovich}]{Abramson2011a}
Abramson, A., Kenney, J. D.~P., Crowl, H.~H., {et~al.} 2011, AJ, 141, 164

\bibitem[{{Arribas} {et~al.}(2023){Arribas}, {Perna}, {Rodr{\'\i}guez Del Pino}, {Lamperti}, {D'Eugenio}, {P{\'e}rez-Gonz{\'a}lez}, {Jones}, {Crespo}, {Curti}, {Bunker}, {Carniani}, {Charlot}, {Jakobsen}, {Maiolino}, {{\"U}bler}, {Willott}, {{\'A}lvarez-M{\'a}rquez}, {B{\"o}ker}, {Chevallard}, {Circosta}, {Cresci}, {Kumari}, {Parlanti}, {Scholtz}, {Venturi}, \& {Witstok}}]{Arribas2023}
{Arribas}, S., {Perna}, M., {Rodr{\'\i}guez Del Pino}, B., {et~al.} 2023, \aap~in press, arXiv:2312.00899

\bibitem[{Athanassoula {et~al.}(1993)Athanassoula, {Garcia Gomez}, \& Bosma}]{Athanassoula1993}
Athanassoula, E., {Garcia Gomez}, C., \& Bosma, A. 1993, A{\&}AS, 102, 229

\bibitem[{{Bagley} {et~al.}(2023){Bagley}, {Finkelstein}, {Koekemoer}, {Ferguson}, {Arrabal Haro}, {Dickinson}, {Kartaltepe}, {Papovich}, {P{\'e}rez-Gonz{\'a}lez}, {Pirzkal}, {Somerville}, {Willmer}, {Yang}, {Yung}, {Fontana}, {Grazian}, {Grogin}, {Hirschmann}, {Kewley}, {Kirkpatrick}, {Kocevski}, {Lotz}, {Medrano}, {Morales}, {Pentericci}, {Ravindranath}, {Trump}, {Wilkins}, {Calabr{\`o}}, {Cooper}, {Costantin}, {de la Vega}, {Hilbert}, {Hutchison}, {Larson}, {Lucas}, {McGrath}, {Ryan}, {Wang}, \& {Wuyts}}]{Bagley2023}
{Bagley}, M.~B., {Finkelstein}, S.~L., {Koekemoer}, A.~M., {et~al.} 2023, \apjl, 946, L12

\bibitem[{Barro {et~al.}(2017)Barro, Faber, Koo, Dekel, Fang, Trump, P{\'{e}}rez-Gonz{\'{a}}lez, Pacifici, Primack, Somerville, Yan, Guo, Liu, Ceverino, Kocevski, \& McGrath}]{Barro2017a}
Barro, G., Faber, S.~M., Koo, D.~C., {et~al.} 2017, ApJ, 840, 47

\bibitem[{Bassett {et~al.}(2017)Bassett, Glazebrook, Fisher, Wisnioski, Damjanov, Abraham, Obreschkow, Green, da~Cunha, \& McGregor}]{Bassett2017}
Bassett, R., Glazebrook, K., Fisher, D.~B., {et~al.} 2017, MNRAS, 467, 239

\bibitem[{{Belfiore} {et~al.}(2017){Belfiore}, {Maiolino}, {Maraston}, {Emsellem}, {Bershady}, {Masters}, {Bizyaev}, {Boquien}, {Brownstein}, {Bundy}, {Diamond-Stanic}, {Drory}, {Heckman}, {Law}, {Malanushenko}, {Oravetz}, {Pan}, {Roman-Lopes}, {Thomas}, {Weijmans}, {Westfall}, \& {Yan}}]{Belfiore2017}
{Belfiore}, F., {Maiolino}, R., {Maraston}, C., {et~al.} 2017, \mnras, 466, 2570

\bibitem[{{Birkin} {et~al.}(2023){Birkin}, {Hutchison}, {Welch}, {Spilker}, {Aravena}, {Bayliss}, {Cathey}, {Chapman}, {Gonzalez}, {Gururajan}, {Hayward}, {Khullar}, {Kim}, {Mahler}, {Malkan}, {Narayanan}, {Olivier}, {Phadke}, {Reuter}, {Rigby}, {Smith}, {Solimano}, {Sulzenauer}, {Vieira}, {Vizgan}, \& {Weiss}}]{Birkin2023}
{Birkin}, J.~E., {Hutchison}, T.~A., {Welch}, B., {et~al.} 2023, \apj, 958, 64

\bibitem[{{B{\"o}ker} {et~al.}(2023){B{\"o}ker}, {Beck}, {Birkmann}, {Giardino}, {Keyes}, {Kumari}, {Muzerolle}, {Rawle}, {Zeidler}, {Abul-Huda}, {Alves de Oliveira}, {Arribas}, {Bechtold}, {Bhatawdekar}, {Bonaventura}, {Bunker}, {Cameron}, {Carniani}, {Charlot}, {Curti}, {Espinoza}, {Ferruit}, {Franx}, {Jakobsen}, {Karakla}, {L{\'o}pez-Caniego}, {L{\"u}tzgendorf}, {Maiolino}, {Manjavacas}, {Marston}, {Moseley}, {Ogle}, {Perna}, {Pe{\~n}a-Guerrero}, {Pirzkal}, {Plesha}, {Proffitt}, {Rauscher}, {Rix}, {Rodr{\'\i}guez del Pino}, {Rustamkulov}, {Sabbi}, {Sing}, {Sirianni}, {te Plate}, {{\'U}beda}, {Wahlgren}, {Wislowski}, {Wu}, \& {Willott}}]{Boker2023}
{B{\"o}ker}, T., {Beck}, T.~L., {Birkmann}, S.~M., {et~al.} 2023, \pasp, 135, 038001

\bibitem[{Boselli {et~al.}(2020)Boselli, Fossati, Longobardi, Boissier, Boquien, Braine, C{\^{o}}t{\'{e}}, Cuillandre, Epinat, Ferrarese, Gavazzi, Gwyn, Hensler, Plana, Roehlly, Schimd, Sun, \& Trinchieri}]{Boselli2020}
Boselli, A., Fossati, M., Longobardi, A., {et~al.} 2020, A{\&}A, 634, L1

\bibitem[{Boselli {et~al.}(2021)Boselli, Lupi, Epinat, Amram, Fossati, Anderson, Boissier, Boquien, Consolandi, C{\^{o}}t{\'{e}}, Cuillandre, Ferrarese, Galbany, Gavazzi, G{\'{o}}mez-L{\'{o}}pez, Gwyn, Hensler, Hutchings, Kuncarayakti, Longobardi, Peng, Plana, Postma, Roediger, Roehlly, Schimd, Trinchieri, \& Vollmer}]{Boselli2021}
Boselli, A., Lupi, A., Epinat, B., {et~al.} 2021, A{\&}A, 646, A139

\bibitem[{{Bouwens} {et~al.}(2015){Bouwens}, {Illingworth}, {Oesch}, {Trenti}, {Labb{\'e}}, {Bradley}, {Carollo}, {van Dokkum}, {Gonzalez}, {Holwerda}, {Franx}, {Spitler}, {Smit}, \& {Magee}}]{Bouwens2015}
{Bouwens}, R.~J., {Illingworth}, G.~D., {Oesch}, P.~A., {et~al.} 2015, \apj, 803, 34

\bibitem[{{Bowler} {et~al.}(2022){Bowler}, {Cullen}, {McLure}, {Dunlop}, \& {Avison}}]{Bowler2022}
{Bowler}, R.~A.~A., {Cullen}, F., {McLure}, R.~J., {Dunlop}, J.~S., \& {Avison}, A. 2022, \mnras, 510, 5088

\bibitem[{Brammer(2016)}]{Brammer2016}
Brammer, G. 2016, Instrum. Sci. Rep. WFC3 2016-16

\bibitem[{Brammer(2022)}]{Grizli2022}
Brammer, G. 2022, grizli, doi:10.5281/zenodo.7351572

\bibitem[{Brammer {et~al.}(2015)Brammer, Ryan, \& Pirzkal}]{Brammer2015}
Brammer, G., Ryan, R., \& Pirzkal, N. 2015, Instrum. Sci. Rep. WFC3 2015-17, 1

\bibitem[{Brammer {et~al.}(2012)Brammer, van Dokkum, Franx, Fumagalli, Patel, Rix, Skelton, Kriek, Nelson, Schmidt, Bezanson, da~Cunha, Erb, Fan, {F{\"{o}}rster Schreiber}, Illingworth, Labb{\'{e}}, Leja, Lundgren, Magee, Marchesini, McCarthy, Momcheva, Muzzin, Quadri, Steidel, Tal, Wake, Whitaker, \& Williams}]{Brammer2012}
Brammer, G.~B., van Dokkum, P.~G., Franx, M., {et~al.} 2012, ApJS, 200, 13

\bibitem[{{Bundy} {et~al.}(2015){Bundy}, {Bershady}, {Law}, {Yan}, {Drory}, {MacDonald}, {Wake}, {Cherinka}, {S{\'a}nchez-Gallego}, {Weijmans}, {Thomas}, {Tremonti}, {Masters}, {Coccato}, {Diamond-Stanic}, {Arag{\'o}n-Salamanca}, {Avila-Reese}, {Badenes}, {Falc{\'o}n-Barroso}, {Belfiore}, {Bizyaev}, {Blanc}, {Bland-Hawthorn}, {Blanton}, {Brownstein}, {Byler}, {Cappellari}, {Conroy}, {Dutton}, {Emsellem}, {Etherington}, {Frinchaboy}, {Fu}, {Gunn}, {Harding}, {Johnston}, {Kauffmann}, {Kinemuchi}, {Klaene}, {Knapen}, {Leauthaud}, {Li}, {Lin}, {Maiolino}, {Malanushenko}, {Malanushenko}, {Mao}, {Maraston}, {McDermid}, {Merrifield}, {Nichol}, {Oravetz}, {Pan}, {Parejko}, {Sanchez}, {Schlegel}, {Simmons}, {Steele}, {Steinmetz}, {Thanjavur}, {Thompson}, {Tinker}, {van den Bosch}, {Westfall}, {Wilkinson}, {Wright}, {Xiao}, \& {Zhang}}]{Bundy2015}
{Bundy}, K., {Bershady}, M.~A., {Law}, D.~R., {et~al.} 2015, \apj, 798, 7

\bibitem[{Calzetti {et~al.}(2000)Calzetti, Armus, Bohlin, Kinney, Koornneef, \& Storchi‐Bergmann}]{Calzetti1999}
Calzetti, D., Armus, L., Bohlin, R.~C., {et~al.} 2000, ApJ, 533, 682

\bibitem[{{Calzetti} {et~al.}(1994){Calzetti}, {Kinney}, \& {Storchi-Bergmann}}]{Calzetti1994}
{Calzetti}, D., {Kinney}, A.~L., \& {Storchi-Bergmann}, T. 1994, \apj, 429, 582

\bibitem[{{Cardelli} {et~al.}(1989){Cardelli}, {Clayton}, \& {Mathis}}]{Cardelli1989}
{Cardelli}, J.~A., {Clayton}, G.~C., \& {Mathis}, J.~S. 1989, \apj, 345, 245

\bibitem[{Cheung {et~al.}(2012)Cheung, Faber, Koo, Dutton, Simard, McGrath, Huang, Bell, Dekel, Fang, Salim, Barro, Bundy, Coil, Cooper, Conselice, Davis, Dom{\'{i}}nguez, Kassin, Kocevski, Koekemoer, Lin, Lotz, Newman, Phillips, Rosario, Weiner, \& Willmer}]{Cheung2012a}
Cheung, E., Faber, S.~M., Koo, D.~C., {et~al.} 2012, ApJ, 760, 131

\bibitem[{Conroy \& Gunn(2010)}]{Conroy2010}
Conroy, C. \& Gunn, J.~E. 2010, ApJ, 712, 833

\bibitem[{Conroy {et~al.}(2009)Conroy, Gunn, \& White}]{Conroy2009}
Conroy, C., Gunn, J.~E., \& White, M. 2009, ApJ, 699, 486

\bibitem[{Cort{\'{e}}s {et~al.}(2006)Cort{\'{e}}s, Kenney, \& Hardy}]{Cortes2006}
Cort{\'{e}}s, J.~R., Kenney, J. D.~P., \& Hardy, E. 2006, AJ, 131, 747

\bibitem[{Cramer {et~al.}(2019)Cramer, Kenney, Sun, Crowl, Yagi, J{\'{a}}chym, Roediger, \& Waldron}]{Cramer2019}
Cramer, W.~J., Kenney, J. D.~P., Sun, M., {et~al.} 2019, ApJ, 870, 63

\bibitem[{Crowl \& Kenney(2006)}]{Crowl2006}
Crowl, H.~H. \& Kenney, J. D.~P. 2006, ApJ, 649, L75

\bibitem[{Dalcanton {et~al.}(1997)Dalcanton, Spergel, \& Summers}]{Dalcanton1997}
Dalcanton, J.~J., Spergel, D.~N., \& Summers, F.~J. 1997, ApJ, 482, 659

\bibitem[{Dekel {et~al.}(2013)Dekel, Zolotov, Tweed, Cacciato, Ceverino, \& Primack}]{Dekel2013}
Dekel, A., Zolotov, A., Tweed, D., {et~al.} 2013, MNRAS, 435, 999

\bibitem[{{D'Eugenio} {et~al.}(2023){D'Eugenio}, {Perez-Gonzalez}, {Maiolino}, {Scholtz}, {Perna}, {Circosta}, {Uebler}, {Arribas}, {Boeker}, {Bunker}, {Carniani}, {Charlot}, {Chevallard}, {Cresci}, {Curtis-Lake}, {Jones}, {Kumari}, {Lamperti}, {Looser}, {Parlanti}, {Rix}, {Robertson}, {Rodriguez Del Pino}, {Tacchella}, {Venturi}, \& {Willott}}]{DEugenio2023}
{D'Eugenio}, F., {Perez-Gonzalez}, P., {Maiolino}, R., {et~al.} 2023, Nature Ast.~in review, arXiv:2308.06317

\bibitem[{{Dunlop} {et~al.}(2021){Dunlop}, {Abraham}, {Ashby}, {Bagley}, {Best}, {Bongiorno}, {Bouwens}, {Bowler}, {Brammer}, {Bremer}, {Calabro'}, {Carnall}, {Castellano}, {Cirasuolo}, {Conselice}, {Cullen}, {Dave}, {Dayal}, {Dekel}, {Dickinson}, {Duncan}, {Elbaz}, {Ellis}, {Ferguson}, {Ferrara}, {Finkelstein}, {Fontana}, {Furlanetto}, {Fynbo}, {Gallerani}, {Gardner}, {Giavalisco}, {Grazian}, {Grogin}, {Harikane}, {Hopkins}, {Ilbert}, {Illingworth}, {Juneau}, {Jung}, {Kartaltepe}, {Kassin}, {Kauffmann}, {Khochfar}, {Kirkpatrick}, {Kocevski}, {Koekemoer}, {Labbe}, {Laporte}, {Larson}, {Lucas}, {Magee}, {Mason}, {McCracken}, {McLeod}, {McLure}, {Merlin}, {Mesinger}, {Milvang-Jensen}, {Newman}, {Oesch}, {Ouchi}, {Pacifici}, {Papovich}, {Peacock}, {Peeples}, {Pentericci}, {Perez-Gonzalez}, {Pirzkal}, {Pope}, {Pye}, {Reddy}, {Robertson}, {Salvato}, {Santini}, {Schaerer}, {Shapley}, {Simons}, {Smit}, {Smith}, {Snyder}, {Somerville}, {Stanway}, {Stefanon}, {Tasca}, {Tikkanen}, {Tresse}, {Trump}, {Whitaker},
  {Wilkins}, {Wright}, {Wyithe}, {van Dokkum}, \& {van der Werf}}]{Dunlop2021}
{Dunlop}, J.~S., {Abraham}, R.~G., {Ashby}, M. L.~N., {et~al.} 2021, {PRIMER: Public Release IMaging for Extragalactic Research}, JWST Proposal. Cycle 1, ID. \#1837

\bibitem[{{Egami} {et~al.}(2023){Egami}, {Sun}, {Alberts}, {Baum}, {Boyett}, {Bunker}, {Cameron}, {Carniani}, {Charlot}, {Chen}, {Chevallard}, {Curti}, {D'Eugenio}, {Danhaive}, {DeCoursey}, {Dudzeviciute}, {Eisenstein}, {Hainline}, {Helton}, {Ji}, {Johnson}, {Kumari}, {Looser}, {Lyu}, {Ma}, {Maiolino}, {Maseda}, {Nelson}, {Rawle}, {Rieke}, {Robertson}, {Sandles}, {Shivaei}, {Smit}, {Suess}, {Tacchella}, {Uebler}, {Whitler}, {Williams}, {Willmer}, {Willott}, {Witstok}, \& {de Graaff}}]{Egami2023}
{Egami}, E., {Sun}, F., {Alberts}, S., {et~al.} 2023, {Complete NIRCam Grism Redshift Survey (CONGRESS)}, JWST Proposal. Cycle 2, ID. \#3577

\bibitem[{Estrada-Carpenter {et~al.}(2019)Estrada-Carpenter, Papovich, Momcheva, Brammer, Long, Quadri, Bridge, Dickinson, Ferguson, Finkelstein, Giavalisco, Gosmeyer, Lotz, Salmon, Skelton, Trump, \& Weiner}]{Estrada-Carpenter2018}
Estrada-Carpenter, V., Papovich, C., Momcheva, I., {et~al.} 2019, ApJ, 870, 133

\bibitem[{Faisst {et~al.}(2018)Faisst, Masters, Wang, Merson, Capak, Malhotra, \& Rhoads}]{Faisst2018}
Faisst, A.~L., Masters, D., Wang, Y., {et~al.} 2018, ApJ, 855, 132

\bibitem[{Fall \& Efstathiou(1980)}]{Fall1980}
Fall, S.~M. \& Efstathiou, G. 1980, MNRAS, 193, 189

\bibitem[{{F{\"{o}}rster Schreiber} {et~al.}(2009){F{\"{o}}rster Schreiber}, Genzel, Bouch{\'{e}}, Cresci, Davies, Buschkamp, Shapiro, Tacconi, Hicks, Genel, Shapley, Erb, Steidel, Lutz, Eisenhauer, Gillessen, Sternberg, Renzini, Cimatti, Daddi, Kurk, Lilly, Kong, Lehnert, Nesvadba, Verma, McCracken, Arimoto, Mignoli, \& Onodera}]{ForsterSchreiber2009}
{F{\"{o}}rster Schreiber}, N.~M., Genzel, R., Bouch{\'{e}}, N., {et~al.} 2009, ApJ, 706, 1364

\bibitem[{{F{\"o}rster Schreiber} {et~al.}(2018){F{\"o}rster Schreiber}, {Renzini}, {Mancini}, {Genzel}, {Bouch{\'e}}, {Cresci}, {Hicks}, {Lilly}, {Peng}, {Burkert}, {Carollo}, {Cimatti}, {Daddi}, {Davies}, {Genel}, {Kurk}, {Lang}, {Lutz}, {Mainieri}, {McCracken}, {Mignoli}, {Naab}, {Oesch}, {Pozzetti}, {Scodeggio}, {Shapiro Griffin}, {Shapley}, {Sternberg}, {Tacchella}, {Tacconi}, {Wuyts}, \& {Zamorani}}]{ForsterSchreiber2018}
{F{\"o}rster Schreiber}, N.~M., {Renzini}, A., {Mancini}, C., {et~al.} 2018, \apjs, 238, 21

\bibitem[{Fossati {et~al.}(2018)Fossati, Mendel, Boselli, Cuillandre, Vollmer, Boissier, Consolandi, Ferrarese, Gwyn, Amram, Boquien, Buat, Burgarella, Cortese, C{\^{o}}t{\'{e}}, C{\^{o}}t{\'{e}}, Durrell, Fumagalli, Gavazzi, Gomez-Lopez, Hensler, Koribalski, Longobardi, Peng, Roediger, Sun, \& Toloba}]{Fossati2018}
Fossati, M., Mendel, J.~T., Boselli, A., {et~al.} 2018, A{\&}A, 614, A57

\bibitem[{{Gardner} {et~al.}(2023){Gardner}, {Mather}, {Abbott}, {Abell}, {Abernathy}, {Abney}, {Abraham}, {Abraham}, {Abul-Huda}, {Acton}, {Adams}, {Adams}, {Adler}, {Adriaensen}, {Aguilar}, {Ahmed}, {Ahmed}, {Ahmed}, {Albat}, {Albert}, {Alberts}, {Aldridge}, {Allen}, {Allen}, {Altenburg}, {Altunc}, {Alvarez}, {{\'A}lvarez-M{\'a}rquez}, {Alves de Oliveira}, {Ambrose}, {Anandakrishnan}, {Andersen}, {Anderson}, {Anderson}, {Anderson}, {Anderson}, {Aprea}, {Archer}, {Arenberg}, {Argyriou}, {Arribas}, {Artigau}, {Arvai}, {Atcheson}, {Atkinson}, {Averbukh}, {Aymergen}, {Bacinski}, {Baggett}, {Bagnasco}, {Baker}, {Balzano}, {Banks}, {Baran}, {Barker}, {Barrett}, {Barringer}, {Barto}, {Bast}, {Baudoz}, {Baum}, {Beatty}, {Beaulieu}, {Bechtold}, {Beck}, {Beddard}, {Beichman}, {Bellagama}, {Bely}, {Berger}, {Bergeron}, {Bernier}, {Bertch}, {Beskow}, {Betz}, {Biagetti}, {Birkmann}, {Bjorklund}, {Blackwood}, {Blazek}, {Blossfeld}, {Bluth}, {Boccaletti}, {Boegner}, {Bohlin}, {Boia}, {B{\"o}ker}, {Bonaventura}, {Bond},
  {Bosley}, {Boucarut}, {Bouchet}, {Bouwman}, {Bower}, {Bowers}, {Bowers}, {Boyce}, {Boyer}, {Boyer}, {Boyer}, {Boyer}, {Bradley}, {Brady}, {Brandl}, {Brannen}, {Breda}, {Bremmer}, {Brennan}, {Bresnahan}, {Bright}, {Broiles}, {Bromenschenkel}, {Brooks}, {Brooks}, {Brown}, {Brown}, {Brown}, {Bruce}, {Bryson}, {Bujanda}, {Bullock}, {Bunker}, {Bureo}, {Burt}, {Bush}, {Bushouse}, {Bussman}, {Cabaud}, {Cale}, {Calhoon}, {Calvani}, {Canipe}, {Caputo}, {Cara}, {Carey}, {Case}, {Cesari}, {Cetorelli}, {Chance}, {Chandler}, {Chaney}, {Chapman}, {Charlot}, {Chayer}, {Cheezum}, {Chen}, {Chen}, {Cherinka}, {Chichester}, {Chilton}, {Chittiraibalan}, {Clampin}, {Clark}, {Clark}, {Clark}, {Claybrooks}, {Cleveland}, {Cohen}, {Cohen}, {Col{\'o}n}, {Coleman}, {Colina}, {Comber}, {Comeau}, {Comer}, {Conde Reis}, {Connolly}, {Conroy}, {Contos}, {Contreras}, {Cook}, {Cooper}, {Cooper}, {Correia}, {Correnti}, {Cossou}, {Costanza}, {Coulais}, {Cox}, {Coyle}, {Cracraft}, {Crew}, {Curtis}, {Cusveller}, {Da Costa Maciel}, {Dailey},
  {Daugeron}, {Davidson}, {Davies}, {Davis}, {Davis}, {Day}, {de Chambure}, {de Jong}, {De Marchi}, {Dean}, {Decker}, {Delisa}, {Dell}, {Dellagatta}, {Dembinska}, {Demosthenes}, {Dencheva}, {Deneu}, {DePriest}, {Deschenes}, {Dethienne}, {Detre}, {Diaz}, {Dicken}, {DiFelice}, {Dillman}, {Disharoon}, {Dixon}, {Doggett}, {Dominguez}, {Donaldson}, {Doria-Warner}, {Santos}, {Doty}, {Douglas}, {Doyon}, {Dressler}, {Driggers}, {Driggers}, {Dunn}, {DuPrie}, {Dupuis}, {Durning}, {Dutta}, {Earl}, {Eccleston}, {Ecobichon}, {Egami}, {Ehrenwinkler}, {Eisenhamer}, {Eisenhower}, {Eisenstein}, {El Hamel}, {Elie}, {Elliott}, {Elliott}, {Engesser}, {Espinoza}, {Etienne}, {Etxaluze}, {Evans}, {Fabreguettes}, {Falcolini}, {Falini}, {Fatig}, {Feeney}, {Feinberg}, {Fels}, {Ferdous}, {Ferguson}, {Ferrarese}, {Ferreira}, {Ferruit}, {Ferry}, {Filippazzo}, {Firre}, {Fix}, {Flagey}, {Flanagan}, {Fleming}, {Florian}, {Flynn}, {Foiadelli}, {Fontaine}, {Fontanella}, {Forshay}, {Fortner}, {Fox}, {Framarini}, {Francisco}, {Franck}, {Franx},
  {Franz}, {Friedman}, {Friend}, {Frost}, {Fu}, {Fullerton}, {Gaillard}, {Galkin}, {Gallagher}, {Galyer}, {Garc{\'\i}a Mar{\'\i}n}, {Gardner}, {Garland}, {Garrett}, {Gasman}, {G{\'a}sp{\'a}r}, {Gastaud}, {Gaudreau}, {Gauthier}, {Geers}, {Geithner}, {Gennaro}, {Gerber}, {Gereau}, {Giampaoli}, {Giardino}, {Gibbons}, {Gilbert}, {Gilman}, {Girard}, {Giuliano}, {Gkountis}, {Glasse}, {Glassmire}, {Glauser}, {Glazer}, {Goldberg}, {Golimowski}, {Gonzaga}, {Gordon}, {Gordon}, {Goudfrooij}, {Gough}, {Graham}, {Grau}, {Green}, {Greene}, {Greene}, {Greenfield}, {Greenhouse}, {Greve}, {Greville}, {Grimaldi}, {Groe}, {Groebner}, {Grumm}, {Grundy}, {G{\"u}del}, {Guillard}, {Guldalian}, {Gunn}, {Gurule}, {Gutman}, {Guy}, {Guyot}, {Hack}, {Haderlein}, {Hagan}, {Hagedorn}, {Hainline}, {Haley}, {Hami}, {Hamilton}, {Hammann}, {Hammel}, {Hanley}, {Hansen}, {Hardy}, {Harnisch}, {Harr}, {Harris}, {Hart}, {Hartig}, {Hasan}, {Hashim}, {Hashimoto}, {Haskins}, {Hawkins}, {Hayden}, {Hayden}, {Healy}, {Hecht}, {Heeg}, {Hejal}, {Helm},
  {Hengemihle}, {Henning}, {Henry}, {Henry}, {Henshaw}, {Hernandez}, {Herrington}, {Heske}, {Hesman}, {Hickey}, {Hilbert}, {Hines}, {Hinz}, {Hirsch}, {Hitcho}, {Hodapp}, {Hodge}, {Hoffman}, {Holfeltz}, {Holler}, {Hoppa}, {Horner}, {Howard}, {Howard}, {Huber}, {Hunkeler}, {Hunter}, {Hunter}, {Hurd}, {Hurst}, {Hutchings}, {Hylan}, {Ignat}, {Illingworth}, {Irish}, {Isaacs}, {Jackson}, {Jaffe}, {Jahic}, {Jahromi}, {Jakobsen}, {James}, {James}, {James}, {Jamieson}, {Jandra}, {Jayawardhana}, {Jedrzejewski}, {Jeffers}, {Jensen}, {Joanne}, {Johns}, {Johnson}, {Johnson}, {Johnson}, {Johnson}, {Johnson}, {Johnson}, {Johnstone}, {Jollet}, {Jones}, {Jones}, {Jones}, {Jones}, {Jones}, {Jordan}, {Jordan}, {Jue}, {Jurkowski}, {Justis}, {Justtanont}, {Kaleida}, {Kalirai}, {Kalmanson}, {Kaltenegger}, {Kammerer}, {Kan}, {Kanarek}, {Kao}, {Karakla}, {Karl}, {Kassin}, {Kauffman}, {Kavanagh}, {Kelley}, {Kelly}, {Kendrew}, {Kennedy}, {Kenny}, {Keski-Kuha}, {Keyes}, {Khan}, {Kidwell}, {Kimble}, {King}, {King}, {Kinzel}, {Kirk},
  {Kirkpatrick}, {Klaassen}, {Klingemann}, {Klintworth}, {Knapp}, {Knight}, {Knollenberg}, {Knutsen}, {Koehler}, {Koekemoer}, {Kofler}, {Kontson}, {Kovacs}, {Kozhurina-Platais}, {Krause}, {Kriss}, {Krist}, {Kristoffersen}, {Krogel}, {Krueger}, {Kulp}, {Kumari}, {Kwan}, {Kyprianou}, {Labador}, {Labiano}, {Lafreni{\`e}re}, {Lagage}, {Laidler}, {Laine}, {Laird}, {Lajoie}, {Lallo}, {Lam}, {LaMassa}, {Lambros}, {Lampenfield}, {Lander}, {Langston}, {Larson}, {Larson}, {LaVerghetta}, {Law}, {Lawrence}, {Lee}, {Lee}, {Lee}, {Leisenring}, {Leveille}, {Levenson}, {Levi}, {Levine}, {Lewis}, {Lewis}, {Lewis}, {Libralato}, {Lidon}, {Liebrecht}, {Lightsey}, {Lilly}, {Lim}, {Lim}, {Ling}, {Link}, {Link}, {Lipinski}, {Liu}, {Lo}, {Lobmeyer}, {Logue}, {Long}, {Long}, {Long}, {Long}, {L{\'o}pez-Caniego}, {Lotz}, {Love-Pruitt}, {Lubskiy}, {Luers}, {Luetgens}, {Luevano}, {Lui}, {Lund}, {Lundquist}, {Lunine}, {L{\"u}tzgendorf}, {Lynch}, {MacDonald}, {MacDonald}, {Macias}, {Macklis}, {Maghami}, {Maharaja}, {Maiolino},
  {Makrygiannis}, {Malla}, {Malumuth}, {Manjavacas}, {Marini}, {Marrione}, {Marston}, {Martel}, {Martin}, {Martin}, {Martinez}, {Maschmann}, {Masci}, {Masetti}, {Maszkiewicz}, {Matthews}, {Matuskey}, {McBrayer}, {McCarthy}, {McCaughrean}, {McClare}, {McClare}, {McCloskey}, {McClurg}, {McCoy}, {McElwain}, {McGregor}, {McGuffey}, {McKay}, {McKenzie}, {McLean}, {McMaster}, {McNeil}, {De Meester}, {Mehalick}, {Meixner}, {Mel{\'e}ndez}, {Menzel}, {Menzel}, {Merz}, {Mesterharm}, {Meyer}, {Meyett}, {Meza}, {Midwinter}, {Milam}, {Miller}, {Miller}, {Miskey}, {Misselt}, {Mitchell}, {Mohan}, {Montoya}, {Moran}, {Morishita}, {Moro-Mart{\'\i}n}, {Morrison}, {Morrison}, {Morse}, {Moschos}, {Moseley}, {Mosier}, {Mosner}, {Mountain}, {Muckenthaler}, {Mueller}, {Mueller}, {Muhiem}, {M{\"u}hlmann}, {Mullally}, {Mullen}, {Munger}, {Murphy}, {Murray}, {Muzerolle}, {Mycroft}, {Myers}, {Myers}, {Myers}, {Myers}, {Myrick}, {Nagle}, {Nayak}, {Naylor}, {Neff}, {Nelan}, {Nella}, {Nguyen}, {Nguyen}, {Nickson}, {Nidhiry}, {Niedner},
  {Nieto-Santisteban}, {Nikolov}, {Nishisaka}, {Noriega-Crespo}, {Nota}, {O'Mara}, {Oboryshko}, {O'Brien}, {Ochs}, {Offenberg}, {Ogle}, {Ohl}, {Olmsted}, {Osborne}, {O'Shaughnessy}, {{\"O}stlin}, {O'Sullivan}, {Otor}, {Ottens}, {Ouellette}, {Outlaw}, {Owens}, {Pacifici}, {Page}, {Paranilam}, {Park}, {Parrish}, {Paschal}, {Patapis}, {Patel}, {Patrick}, {Pattishall}, {Paul}, {Paul}, {Pauly}, {Pavlovsky}, {Pe{\~n}a-Guerrero}, {Pedder}, {Peek}, {Pelham}, {Penanen}, {Perriello}, {Perrin}, {Perrine}, {Perrygo}, {Peslier}, {Petach}, {Peterson}, {Pfarr}, {Pierson}, {Pietraszkiewicz}, {Pilchen}, {Pipher}, {Pirzkal}, {Pitman}, {Player}, {Plesha}, {Plitzke}, {Pohner}, {Poletis}, {Pollizzi}, {Polster}, {Pontius}, {Pontoppidan}, {Porges}, {Potter}, {Prescott}, {Proffitt}, {Pueyo}, {Quispe Neira}, {Radich}, {Rager}, {Rameau}, {Ramey}, {Ramos Alarcon}, {Rampini}, {Rapp}, {Rashford}, {Rauscher}, {Ravindranath}, {Rawle}, {Rawlings}, {Ray}, {Regan}, {Rehm}, {Rehm}, {Reid}, {Reis}, {Renk}, {Reoch}, {Ressler}, {Rest},
  {Reynolds}, {Richon}, {Richon}, {Ridgaway}, {Riedel}, {Rieke}, {Rieke}, {Rifelli}, {Rigby}, {Riggs}, {Ringel}, {Ritchie}, {Rix}, {Robberto}, {Robinson}, {Robinson}, {Robinson}, {Rock}, {Rodriguez}, {Rodr{\'\i}guez del Pino}, {Roellig}, {Rohrbach}, {Roman}, {Romelfanger}, {Romo}, {Rosales}, {Rose}, {Roteliuk}, {Roth}, {Rothwell}, {Rouzaud}, {Rowe}, {Rowlands}, {Roy}, {Royer}, {Rui}, {Rumler}, {Rumpl}, {Russ}, {Ryan}, {Ryan}, {Saad}, {Sabata}, {Sabatino}, {Sabbi}, {Sabelhaus}, {Sabia}, {Sahu}, {Saif}, {Salvignol}, {Samara-Ratna}, {Samuelson}, {Sanders}, {Sappington}, {Sargent}, {Sauer}, {Savadkin}, {Sawicki}, {Schappell}, {Scheffer}, {Scheithauer}, {Scherer}, {Schiff}, {Schlawin}, {Schmeitzky}, {Schmitz}, {Schmude}, {Schneider}, {Schreiber}, {Schroeven-Deceuninck}, {Schultz}, {Schwab}, {Schwartz}, {Scoccimarro}, {Scott}, {Scott}, {Seaton}, {Seely}, {Seery}, {Seidleck}, {Sembach}, {Shanahan}, {Shaughnessy}, {Shaw}, {Shay}, {Sheehan}, {Sheth}, {Shih}, {Shivaei}, {Siegel}, {Sienkiewicz}, {Simmons}, {Simon},
  {Sirianni}, {Sivaramakrishnan}, {Slade}, {Sloan}, {Slocum}, {Slowinski}, {Smith}, {Smith}, {Smith}, {Smith}, {Smith}, {Smith}, {Smolik}, {Soderblom}, {Sohn}, {Sokol}, {Sonneborn}, {Sontag}, {Sooy}, {Soummer}, {Southwood}, {Spain}, {Sparmo}, {Speer}, {Spencer}, {Sprofera}, {Stallcup}, {Stanley}, {Stansberry}, {Stark}, {Starr}, {Stassi}, {Steck}, {Steeley}, {Stephens}, {Stephenson}, {Stewart}, {Stiavelli}, {}, {Strada}, {Straughn}, {Streetman}, {Strickland}, {Strobele}, {Stuhlinger}, {Stys}, {Such}, {Sukhatme}, {Sullivan}, {Sullivan}, {Sumner}, {Sun}, {Sunnquist}, {Swade}, {Swam}, {Swenton}, {Swoish}, {Tam Litten}, {Tamas}, {Tao}, {Taylor}, {Taylor}, {te Plate}, {Van Tea}, {Teague}, {Telfer}, {Temim}, {Texter}, {Thatte}, {Thompson}, {Thompson}, {Thomson}, {Thronson}, {Tierney}, {Tikkanen}, {Tinnin}, {Tippet}, {Todd}, {Tran}, {Trauger}, {Trejo}, {Vinh Truong}, {Tsukamoto}, {Tufail}, {Tumlinson}, {Tustain}, {Tyra}, {Ubeda}, {Underwood}, {Uzzo}, {Vaclavik}, {Valenduc}, {Valenti}, {Van Campen}, {van de Wetering},
  {Van Der Marel}, {van Haarlem}, {Vandenbussche}, {van Dishoeck}, {Vanterpool}, {Vernoy}, {Vila Costas}, {Volk}, {Voorzaat}, {Voyton}, {Vydra}, {Waddy}, {Waelkens}, {Wahlgren}, {Walker}, {Wander}, {Warfield}, {Warner}, {Wasiak}, {Wasiak}, {Wehner}, {Weiler}, {Weilert}, {Weiss}, {Wells}, {Welty}, {Wheate}, {Wheeler}, {White}, {Whitehouse}, {Whiteleather}, {Whitman}, {Williams}, {Willmer}, {Willott}, {Willoughby}, {Wilson}, {Wilson}, {Wilson}, {Windhorst}, {Wislowski}, {Wolfe}, {Wolfe}, {Wolff}, {Wondel}, {Woo}, {Woods}, {Worden}, {Workman}, {Wright}, {Wu}, {Wu}, {Wun}, {Wymer}, {Yadetie}, {Yan}, {Yang}, {Yates}, {Yeager}, {Yerger}, {Young}, {Young}, {Yu}, {Yu}, {Zak}, {Zeidler}, {Zepp}, {Zhou}, {Zincke}, {Zonak}, \& {Zondag}}]{Gardner2023}
{Gardner}, J.~P., {Mather}, J.~C., {Abbott}, R., {et~al.} 2023, \pasp, 135, 068001

\bibitem[{{Gavazzi} {et~al.}(2018){Gavazzi}, {Consolandi}, {Gutierrez}, {Boselli}, \& {Yoshida}}]{Gavazzi2018}
{Gavazzi}, G., {Consolandi}, G., {Gutierrez}, M.~L., {Boselli}, A., \& {Yoshida}, M. 2018, \aap, 618, A130

\bibitem[{Gavazzi {et~al.}(2013)Gavazzi, Fumagalli, Fossati, Galardo, Grossetti, Boselli, Giovanelli, \& Haynes}]{Gavazzi2013}
Gavazzi, G., Fumagalli, M., Fossati, M., {et~al.} 2013, A{\&}A, 553, 1

\bibitem[{{George} {et~al.}(2024){George}, {Damjanov}, {Sawicki}, {Arnouts}, {Desprez}, {Gwyn}, {Picouet}, {Birrer}, \& {Silverman}}]{George2024}
{George}, A., {Damjanov}, I., {Sawicki}, M., {et~al.} 2024, \mnras, 528, 4797

\bibitem[{{Gnerucci} {et~al.}(2011){Gnerucci}, {Marconi}, {Cresci}, {Maiolino}, {Mannucci}, {Calura}, {Cimatti}, {Cocchia}, {Grazian}, {Matteucci}, {Nagao}, {Pozzetti}, \& {Troncoso}}]{Gnerucci2011}
{Gnerucci}, A., {Marconi}, A., {Cresci}, G., {et~al.} 2011, \aap, 528, A88

\bibitem[{Greener {et~al.}(2020)Greener, Arag{\'{o}}n-Salamanca, Merrifield, Peterken, Fraser-McKelvie, Masters, Krawczyk, Boardman, Boquien, Andrews, Brinkmann, \& Drory}]{Greener2020}
Greener, M.~J., Arag{\'{o}}n-Salamanca, A., Merrifield, M.~R., {et~al.} 2020, MNRAS, 495, 2305

\bibitem[{Hodge \& {Kennicutt, R. C.}(1983)}]{Hodge1983}
Hodge, P.~W. \& {Kennicutt, R. C.}, J. 1983, ApJ, 267, 563

\bibitem[{{Hutchison} {et~al.}(2024){Hutchison}, {Welch}, {Rigby}, {Olivier}, {Birkin}, {Phadke}, {Khullar}, {Rauscher}, {Sharon}, {Aravena}, {Bayliss}, {Elicker}, {Kim}, {Solimano}, {Vieira}, {Vizgan}, \& {Jwst Templates Early Release Science Team}}]{Hutchison2023}
{Hutchison}, T.~A., {Welch}, B.~D., {Rigby}, J.~R., {et~al.} 2024, \pasp, 136, 044503

\bibitem[{{Johnson} {et~al.}(2021){Johnson}, {Leja}, {Conroy}, \& {Speagle}}]{Johnson2021}
{Johnson}, B.~D., {Leja}, J., {Conroy}, C., \& {Speagle}, J.~S. 2021, \apjs, 254, 22

\bibitem[{{Johnson} {et~al.}(2018){Johnson}, {Harrison}, {Swinbank}, {Tiley}, {Stott}, {Bower}, {Smail}, {Bunker}, {Sobral}, {Turner}, {Best}, {Bureau}, {Cirasuolo}, {Jarvis}, {Magdis}, {Sharples}, {Bland-Hawthorn}, {Catinella}, {Cortese}, {Croom}, {Federrath}, {Glazebrook}, {Sweet}, {Bryant}, {Goodwin}, {Konstantopoulos}, {Lawrence}, {Medling}, {Owers}, \& {Richards}}]{Johnson2018}
{Johnson}, H.~L., {Harrison}, C.~M., {Swinbank}, A.~M., {et~al.} 2018, \mnras, 474, 5076

\bibitem[{{Jones} {et~al.}(2024){Jones}, {{\"U}bler}, {Perna}, {Arribas}, {Bunker}, {Carniani}, {Charlot}, {Maiolino}, {Del Pino}, {Willott}, {Bowler}, {B{\"o}ker}, {Cameron}, {Chevallard}, {Cresci}, {Curti}, {D'Eugenio}, {Kumari}, {Saxena}, {Scholtz}, {Venturi}, \& {Witstok}}]{Jones2024}
{Jones}, G.~C., {{\"U}bler}, H., {Perna}, M., {et~al.} 2024, \aap, 682, A122

\bibitem[{{Kashino} {et~al.}(2023){Kashino}, {Lilly}, {Matthee}, {Eilers}, {Mackenzie}, {Bordoloi}, \& {Simcoe}}]{Kashino2023}
{Kashino}, D., {Lilly}, S.~J., {Matthee}, J., {et~al.} 2023, \apj, 950, 66

\bibitem[{Kashino {et~al.}(2013)Kashino, Silverman, Rodighiero, Renzini, Arimoto, Daddi, Lilly, Sanders, Kartaltepe, Zahid, Nagao, Sugiyama, Capak, Carollo, Chu, Hasinger, Ilbert, Kajisawa, Kewley, Koekemoer, Kova{\v{c}}, {Le F{\`{e}}vre}, Masters, McCracken, Onodera, Scoville, Strazzullo, Symeonidis, \& Taniguchi}]{Kashino2013}
Kashino, D., Silverman, J.~D., Rodighiero, G., {et~al.} 2013, ApJL, 777, 4

\bibitem[{{Kassin} {et~al.}(2012){Kassin}, {Weiner}, {Faber}, {Gardner}, {Willmer}, {Coil}, {Cooper}, {Devriendt}, {Dutton}, {Guhathakurta}, {Koo}, {Metevier}, {Noeske}, \& {Primack}}]{Kassin2012}
{Kassin}, S.~A., {Weiner}, B.~J., {Faber}, S.~M., {et~al.} 2012, \apj, 758, 106

\bibitem[{Kenney {et~al.}(2015)Kenney, Abramson, \& Bravo-Alfaro}]{Kenney2015}
Kenney, J.~D., Abramson, A., \& Bravo-Alfaro, H. 2015, AJ, 150, 59

\bibitem[{Kenney \& Koopmann(1999)}]{Kenney1999}
Kenney, J. D.~P. \& Koopmann, R.~A. 1999, AJ, 117, 181

\bibitem[{Kennicutt(1998)}]{Kennicutt1998}
Kennicutt, R.~C. 1998, ARAA, 36, 189

\bibitem[{Kennicutt \& Evans(2012)}]{Kennicutt2012}
Kennicutt, R.~C. \& Evans, N.~J. 2012, ARAA, 50, 531

\bibitem[{Koopmann {et~al.}(2006)Koopmann, Haynes, \& Catinella}]{Koopmann2006}
Koopmann, R.~A., Haynes, M.~P., \& Catinella, B. 2006, AJ, 131, 716

\bibitem[{Koopmann \& Kenney(2004{\natexlab{a}})}]{Koopmann2004a}
Koopmann, R.~a. \& Kenney, J. D.~P. 2004{\natexlab{a}}, ApJ, 613, 866

\bibitem[{Koopmann \& Kenney(2004{\natexlab{b}})}]{Koopmann2004}
Koopmann, R.~A. \& Kenney, J. D.~P. 2004{\natexlab{b}}, ApJ, 613, 851

\bibitem[{Koyama {et~al.}(2019)Koyama, Shimakawa, Yamamura, Kodama, \& Hayashi}]{Koyama2019a}
Koyama, Y., Shimakawa, R., Yamamura, I., Kodama, T., \& Hayashi, M. 2019, PASJ, 71, 1

\bibitem[{{Kreckel} {et~al.}(2013){Kreckel}, {Groves}, {Schinnerer}, {Johnson}, {Aniano}, {Calzetti}, {Croxall}, {Draine}, {Gordon}, {Crocker}, {Dale}, {Hunt}, {Kennicutt}, {Meidt}, {Smith}, \& {Tabatabaei}}]{Kreckel2013}
{Kreckel}, K., {Groves}, B., {Schinnerer}, E., {et~al.} 2013, \apj, 771, 62

\bibitem[{{Kuhn} {et~al.}(2024){Kuhn}, {Guo}, {Martin}, {Bayless}, {Gates}, \& {Puleo}}]{Kuhn2023}
{Kuhn}, V., {Guo}, Y., {Martin}, A., {et~al.} 2024, \apjl, 968, L15

\bibitem[{Lee {et~al.}(2017)Lee, Chung, Tonnesen, Kenney, {Ivy Wong}, Vollmer, Petitpas, Crowl, \& van Gorkom}]{Lee2017}
Lee, B., Chung, A., Tonnesen, S., {et~al.} 2017, MNRAS, 466, 1382

\bibitem[{{Leja} {et~al.}(2019){Leja}, {Carnall}, {Johnson}, {Conroy}, \& {Speagle}}]{Leja2019}
{Leja}, J., {Carnall}, A.~C., {Johnson}, B.~D., {Conroy}, C., \& {Speagle}, J.~S. 2019, \apj, 876, 3

\bibitem[{{Leja} {et~al.}(2017){Leja}, {Johnson}, {Conroy}, {van Dokkum}, \& {Byler}}]{Leja2017}
{Leja}, J., {Johnson}, B.~D., {Conroy}, C., {van Dokkum}, P.~G., \& {Byler}, N. 2017, \apj, 837, 170

\bibitem[{{Li} {et~al.}(2023){Li}, {Cai}, {Sun}, {Richard}, {Trebitsch}, {Helton}, {Diego}, {Oguri}, {Foo}, {Lin}, {Bauer}, {Chen}, {Conselice}, {Espada}, {Egami}, {Fan}, {Frye}, {Fudamoto}, {Perez-Gonzalez}, {Hainline}, {Hsiao}, {Ji}, {Jin}, {Koekemoer}, {Kokorev}, {Kohno}, {Li}, {Lee}, {Magdis}, {Willmer}, {Windhorst}, {Wu}, {Yan}, {Zhang}, {Zitrin}, {Zou}, {Bian}, {Cheng}, {DeCoursey}, {Furtak}, {Steinhardt}, \& {Umehata}}]{Zihao2023}
{Li}, Z., {Cai}, Z., {Sun}, F., {et~al.} 2023, \apj~in review, arXiv:2310.09327

\bibitem[{{Liu} {et~al.}(2023){Liu}, {Morishita}, \& {Kodama}}]{Liu2023}
{Liu}, Z., {Morishita}, T., \& {Kodama}, T. 2023, \apj, 955, 29

\bibitem[{{Loiacono} {et~al.}(2024){Loiacono}, {Decarli}, {Mignoli}, {Farina}, {Ba{\~n}ados}, {Bosman}, {Eilers}, {Schindler}, {Strauss}, {Vestergaard}, {Wang}, {Blecha}, {Carilli}, {Comastri}, {Connor}, {Costa}, {Dotti}, {Fan}, {Gilli}, {Jun}, {Liu}, {Lupi}, {Marshall}, {Mazzucchelli}, {Meyer}, {Neeleman}, {Overzier}, {Pensabene}, {Riechers}, {Trakhtenbrot}, {Trebitsch}, {Venemans}, {Walter}, \& {Yang}}]{Loiacono2024}
{Loiacono}, F., {Decarli}, R., {Mignoli}, M., {et~al.} 2024, \aap, 685, A121

\bibitem[{{Mancini} {et~al.}(2011){Mancini}, {F{\"o}rster Schreiber}, {Renzini}, {Cresci}, {Hicks}, {Peng}, {Vergani}, {Lilly}, {Carollo}, {Pozzetti}, {Zamorani}, {Daddi}, {Genzel}, {Maraston}, {McCracken}, {Tacconi}, {Bouch{\'e}}, {Davies}, {Oesch}, {Shapiro}, {Mainieri}, {Lutz}, {Mignoli}, \& {Sternberg}}]{Mancini2011}
{Mancini}, C., {F{\"o}rster Schreiber}, N.~M., {Renzini}, A., {et~al.} 2011, \apj, 743, 86

\bibitem[{{Matharu} {et~al.}(2021){Matharu}, {Muzzin}, {Brammer}, {Nelson}, {Auger}, {Hewett}, {van der Burg}, {Balogh}, {Demarco}, {Marchesini}, {Noble}, {Rudnick}, {van der Wel}, {Wilson}, \& {Yee}}]{Matharu2021}
{Matharu}, J., {Muzzin}, A., {Brammer}, G.~B., {et~al.} 2021, \apj, 923, 222

\bibitem[{Matharu {et~al.}(2019)Matharu, Muzzin, Brammer, van~der Burg, Auger, Hewett, van~der Wel, van Dokkum, Balogh, Chan, Demarco, Marchesini, Nelson, Noble, Wilson, \& Yee}]{Matharu2018}
Matharu, J., Muzzin, A., Brammer, G.~B., {et~al.} 2019, MNRAS, 484, 595

\bibitem[{{Matharu} {et~al.}(2023){Matharu}, {Muzzin}, {Sarrouh}, {Brammer}, {Abraham}, {Asada}, {Brada{\v{c}}}, {Desprez}, {Martis}, {Mowla}, {Noirot}, {Sawicki}, {Strait}, {Willott}, {Gould}, {Grindlay}, \& {Harshan}}]{Matharu2023}
{Matharu}, J., {Muzzin}, A., {Sarrouh}, G. T.~E., {et~al.} 2023, \apjl, 949, L11

\bibitem[{{Matharu} {et~al.}(2022){Matharu}, {Papovich}, {Simons}, {Momcheva}, {Brammer}, {Ji}, {Backhaus}, {Cleri}, {Estrada-Carpenter}, {Finkelstein}, {Finlator}, {Giavalisco}, {Jung}, {Muzzin}, {Nelson}, {Pillepich}, {Trump}, \& {Weiner}}]{Matharu2022}
{Matharu}, J., {Papovich}, C., {Simons}, R.~C., {et~al.} 2022, \apj, 937, 16

\bibitem[{{Miller} {et~al.}(2012){Miller}, {Ellis}, {Sullivan}, {Bundy}, {Newman}, \& {Treu}}]{Miller2012}
{Miller}, S.~H., {Ellis}, R.~S., {Sullivan}, M., {et~al.} 2012, \apj, 753, 74

\bibitem[{{Mitsuhashi} {et~al.}(2023){Mitsuhashi}, {Tadaki}, {Ikeda}, {Herrera-Camus}, {Aravena}, {De Looze}, {F{\"o}rster Schreiber}, {Gonz{\'a}lez-L{\'o}pez}, {Spilker}, {Assef}, {Bouwens}, {Barcos-Munoz}, {Birkin}, {Bowler}, {Calistro Rivera}, {Davies}, {Da Cunha}, {D{\'\i}az-Santos}, {Ferrara}, {Fisher}, {Lee}, {Li}, {Lutz}, {Rela{\~n}o}, {Naab}, {Palla}, {Posses}, {Solimano}, {Tacconi}, {{\"U}bler}, {van der Giessen}, \& {Veilleux}}]{Mitsuhashi2023}
{Mitsuhashi}, I., {Tadaki}, K.-i., {Ikeda}, R., {et~al.} 2023, \aap~in review, arXiv:2311.17671

\bibitem[{Mo {et~al.}(1998)Mo, Mao, \& White}]{Mo1998}
Mo, H.~J., Mao, S., \& White, S. D.~M. 1998, MNRAS, 295, 319

\bibitem[{Momcheva {et~al.}(2016)Momcheva, Brammer, van Dokkum, Skelton, Whitaker, Nelson, Fumagalli, Maseda, Leja, Franx, Rix, Bezanson, Cunha, Dickey, Schreiber, Illingworth, Kriek, Labb{\'{e}}, Lange, Lundgren, Magee, Marchesini, Oesch, Pacifici, Patel, Price, Tal, Wake, van~der Wel, \& Wuyts}]{Momcheva2016}
Momcheva, I.~G., Brammer, G.~B., van Dokkum, P.~G., {et~al.} 2016, ApJS, 225, 27

\bibitem[{{Morishita} {et~al.}(2024){Morishita}, {Stiavelli}, {Chary}, {Trenti}, {Bergamini}, {Chiaberge}, {Leethochawalit}, {Roberts-Borsani}, {Shen}, \& {Treu}}]{Morishita2023}
{Morishita}, T., {Stiavelli}, M., {Chary}, R.-R., {et~al.} 2024, \apj, 963, 9

\bibitem[{{Mowla} {et~al.}(2019){Mowla}, {van Dokkum}, {Brammer}, {Momcheva}, {van der Wel}, {Whitaker}, {Nelson}, {Bezanson}, {Muzzin}, {Franx}, {MacKenty}, {Leja}, {Kriek}, \& {Marchesini}}]{Mowla2019}
{Mowla}, L.~A., {van Dokkum}, P., {Brammer}, G.~B., {et~al.} 2019, \apj, 880, 57

\bibitem[{Munoz‐Mateos {et~al.}(2007)Munoz‐Mateos, {Gil de Paz}, Boissier, Zamorano, Jarrett, Gallego, \& Madore}]{MunozMateos2007}
Munoz‐Mateos, J.~C., {Gil de Paz}, A., Boissier, S., {et~al.} 2007, ApJ, 658, 1006

\bibitem[{{Naidu} {et~al.}(2022){Naidu}, {Oesch}, {van Dokkum}, {Nelson}, {Suess}, {Brammer}, {Whitaker}, {Illingworth}, {Bouwens}, {Tacchella}, {Matthee}, {Allen}, {Bezanson}, {Conroy}, {Labbe}, {Leja}, {Leonova}, {Magee}, {Price}, {Setton}, {Strait}, {Stefanon}, {Toft}, {Weaver}, \& {Weibel}}]{Naidu2022}
{Naidu}, R.~P., {Oesch}, P.~A., {van Dokkum}, P., {et~al.} 2022, \apjl, 940, L14

\bibitem[{{Nedkova} {et~al.}(2021){Nedkova}, {H{\"a}u{\ss}ler}, {Marchesini}, {Dimauro}, {Brammer}, {Eigenthaler}, {Feinstein}, {Ferguson}, {Huertas-Company}, {Johnston}, {Kado-Fong}, {Kartaltepe}, {Labb{\'e}}, {Lange-Vagle}, {Martis}, {McGrath}, {Muzzin}, {Oesch}, {Ordenes-Brice{\~n}o}, {Puzia}, {Shipley}, {Simmons}, {Skelton}, {Stefanon}, {van der Wel}, \& {Whitaker}}]{Nedkova2021}
{Nedkova}, K.~V., {H{\"a}u{\ss}ler}, B., {Marchesini}, D., {et~al.} 2021, \mnras, 506, 928

\bibitem[{{Nelson} {et~al.}(2023){Nelson}, {Brammer}, {Gimenez-Arteaga}, {Oesch}, {Ubler}, {de Graaff}, {Matharu}, {Naidu}, {Shapley}, {Whitaker}, {Wisnioski}, {Forster Schreiber}, {Smit}, {van Dokkum}, {Chisholm}, {Endsley}, {Hartley}, {Gibson}, {Giovinazzo}, {Illingworth}, {Labbe}, {Maseda}, {Matthee}, {Covelo Paz}, {Price}, {Reddy}, {Shivaei}, {Weibel}, {Wuyts}, {Xiao}, {Alberts}, {Baker}, {Bunker}, {Cameron}, {Charlot}, {Eisenstein}, {Ji}, {Johnson}, {Jones}, {Maiolino}, {Robertson}, {Sandles}, {Suess}, {Tacchella}, {Williams}, \& {Witstok}}]{Nelson2023}
{Nelson}, E.~J., {Brammer}, G., {Gimenez-Arteaga}, C., {et~al.} 2023, \apj~in review, arXiv:2310.06887

\bibitem[{Nelson {et~al.}(2012)Nelson, {Van Dokkum}, Brammer, {F{\"{o}}rster Schreiber}, Franx, Fumagalli, Patel, Rix, Skelton, Bezanson, {Da Cunha}, Kriek, Labbe, Lundgren, Quadri, \& Schmidt}]{Nelson2012}
Nelson, E.~J., {Van Dokkum}, P.~G., Brammer, G., {et~al.} 2012, ApJL, 747, 6

\bibitem[{Nelson {et~al.}(2016{\natexlab{a}})Nelson, van Dokkum, {F{\"{o}}rster Schreiber}, Franx, Brammer, Momcheva, Wuyts, Whitaker, Skelton, Fumagalli, Hayward, Kriek, Labb{\'{e}}, Leja, Rix, Tacconi, van~der Wel, van~den Bosch, Oesch, Dickey, \& Lange}]{Nelson2016}
Nelson, E.~J., van Dokkum, P.~G., {F{\"{o}}rster Schreiber}, N.~M., {et~al.} 2016{\natexlab{a}}, ApJ, 828, 27

\bibitem[{Nelson {et~al.}(2016{\natexlab{b}})Nelson, van Dokkum, Momcheva, Brammer, Wuyts, Franx, Schreiber, Whitaker, \& Skelton}]{Nelson2016a}
Nelson, E.~J., van Dokkum, P.~G., Momcheva, I.~G., {et~al.} 2016{\natexlab{b}}, ApJ, 817, L9

\bibitem[{{Noirot} {et~al.}(2022){Noirot}, {Sawicki}, {Abraham}, {Brada{\v{c}}}, {Iyer}, {Moutard}, {Pacifici}, {Ravindranath}, \& {Willott}}]{Noirot2022}
{Noirot}, G., {Sawicki}, M., {Abraham}, R., {et~al.} 2022, \mnras, 512, 3566

\bibitem[{{Oesch} {et~al.}(2023){Oesch}, {Brammer}, {Naidu}, {Bouwens}, {Chisholm}, {Illingworth}, {Matthee}, {Nelson}, {Qin}, {Reddy}, {Shapley}, {Shivaei}, {van Dokkum}, {Weibel}, {Whitaker}, {Wuyts}, {Covelo-Paz}, {Endsley}, {Fudamoto}, {Giovinazzo}, {Herard-Demanche}, {Kerutt}, {Kramarenko}, {Labbe}, {Leonova}, {Lin}, {Magee}, {Marchesini}, {Maseda}, {Mason}, {Matharu}, {Meyer}, {Neufeld}, {Prieto Lyon}, {Schaerer}, {Sharma}, {Shuntov}, {Smit}, {Stefanon}, {Wyithe}, \& {Xiao}}]{Oesch2023}
{Oesch}, P.~A., {Brammer}, G., {Naidu}, R.~P., {et~al.} 2023, \mnras, 525, 2864

\bibitem[{{Parlanti} {et~al.}(2024){Parlanti}, {Carniani}, {{\"U}bler}, {Venturi}, {Circosta}, {D'Eugenio}, {Arribas}, {Bunker}, {Charlot}, {L{\"u}tzgendorf}, {Maiolino}, {Perna}, {Rodr{\'\i}guez Del Pino}, {Willott}, {B{\"o}ker}, {Cameron}, {Chevallard}, {Cresci}, {Jones}, {Kumari}, {Lamperti}, \& {Scholtz}}]{Parlanti2023}
{Parlanti}, E., {Carniani}, S., {{\"U}bler}, H., {et~al.} 2024, \aap, 684, A24

\bibitem[{Peng {et~al.}(2010)Peng, Ho, Impey, \& Rix}]{PengGALFIT2010}
Peng, C., Ho, L., Impey, C., \& Rix, H.-W. 2010, 139, 2097

\bibitem[{Peng {et~al.}(2002)Peng, Ho, Impey, \& Rix}]{Peng2002}
Peng, C.~Y., Ho, L.~C., Impey, C.~D., \& Rix, H.-W. 2002, AJ, 124, 266

\bibitem[{{Perna} {et~al.}(2023){Perna}, {Arribas}, {Marshall}, {D'Eugenio}, {{\"U}bler}, {Bunker}, {Charlot}, {Carniani}, {Jakobsen}, {Maiolino}, {Rodr{\'\i}guez Del Pino}, {Willott}, {B{\"o}ker}, {Circosta}, {Cresci}, {Curti}, {Husemann}, {Kumari}, {Lamperti}, {P{\'e}rez-Gonz{\'a}lez}, \& {Scholtz}}]{Perna2023}
{Perna}, M., {Arribas}, S., {Marshall}, M., {et~al.} 2023, \aap, 679, A89

\bibitem[{{Planck Collaboration XIII}(2016)}]{Planck2015}
{Planck Collaboration XIII}. 2016, A{\&}A, 594, A13

\bibitem[{{Popesso} {et~al.}(2023){Popesso}, {Concas}, {Cresci}, {Belli}, {Rodighiero}, {Inami}, {Dickinson}, {Ilbert}, {Pannella}, \& {Elbaz}}]{Popesso2023}
{Popesso}, P., {Concas}, A., {Cresci}, G., {et~al.} 2023, \mnras, 519, 1526

\bibitem[{{Price} {et~al.}(2020){Price}, {Kriek}, {Barro}, {Shapley}, {Reddy}, {Freeman}, {Coil}, {Shivaei}, {Azadi}, {de Groot}, {Siana}, {Mobasher}, {Sanders}, {Leung}, {Fetherolf}, {Zick}, {{\"U}bler}, \& {F{\"o}rster Schreiber}}]{Price2020}
{Price}, S.~H., {Kriek}, M., {Barro}, G., {et~al.} 2020, \apj, 894, 91

\bibitem[{{Price} {et~al.}(2014){Price}, {Kriek}, {Brammer}, {Conroy}, {F{\"o}rster Schreiber}, {Franx}, {Fumagalli}, {Lundgren}, {Momcheva}, {Nelson}, {Skelton}, {van Dokkum}, {Whitaker}, \& {Wuyts}}]{Price2014}
{Price}, S.~H., {Kriek}, M., {Brammer}, G.~B., {et~al.} 2014, \apj, 788, 86

\bibitem[{Reddy {et~al.}(2015)Reddy, Kriek, Shapley, Freeman, Siana, Coil, Mobasher, Price, Sanders, \& Shivaei}]{Reddy2015}
Reddy, N.~A., Kriek, M., Shapley, A.~E., {et~al.} 2015, ApJ, 806, 259

\bibitem[{{Rieke} {et~al.}(2003){Rieke}, {Baum}, {Beichman}, {Crampton}, {Doyon}, {Eisenstein}, {Greene}, {Hodapp}, {Horner}, {Johnstone}, {Lesyna}, {Lilly}, {Meyer}, {Martin}, {McCarthy}, {Rieke}, {Roellig}, {Stauffer}, {Trauger}, \& {Young}}]{Rieke2003}
{Rieke}, M.~J., {Baum}, S.~A., {Beichman}, C.~A., {et~al.} 2003, in Society of Photo-Optical Instrumentation Engineers (SPIE) Conference Series, Vol. 4850, IR Space Telescopes and Instruments, ed. J.~C. {Mather}, 478--485

\bibitem[{{Rieke} {et~al.}(2005){Rieke}, {Kelly}, \& {Horner}}]{Rieke2005}
{Rieke}, M.~J., {Kelly}, D., \& {Horner}, S. 2005, in Society of Photo-Optical Instrumentation Engineers (SPIE) Conference Series, Vol. 5904, Cryogenic Optical Systems and Instruments XI, ed. J.~B. {Heaney} \& L.~G. {Burriesci}, 1--8

\bibitem[{{Rigby} {et~al.}(2023{\natexlab{a}}){Rigby}, {Perrin}, {McElwain}, {Kimble}, {Friedman}, {Lallo}, {Doyon}, {Feinberg}, {Ferruit}, {Glasse}, {Rieke}, {Rieke}, {Wright}, {Willott}, {Colon}, {Milam}, {Neff}, {Stark}, {Valenti}, {Abell}, {Abney}, {Abul-Huda}, {Acton}, {Adams}, {Adler}, {Aguilar}, {Ahmed}, {Albert}, {Alberts}, {Aldridge}, {Allen}, {Altenburg}, {{\'A}lvarez-M{\'a}rquez}, {Alves de Oliveira}, {Andersen}, {Anderson}, {Anderson}, {Argyriou}, {Armstrong}, {Arribas}, {Artigau}, {Arvai}, {Atkinson}, {Bacon}, {Bair}, {Banks}, {Barrientes}, {Barringer}, {Bartosik}, {Bast}, {Baudoz}, {Beatty}, {Bechtold}, {Beck}, {Bergeron}, {Bergkoetter}, {Bhatawdekar}, {Birkmann}, {Blazek}, {Blome}, {Boccaletti}, {B{\"o}ker}, {Boia}, {Bonaventura}, {Bond}, {Bosley}, {Boucarut}, {Bourque}, {Bouwman}, {Bower}, {Bowers}, {Boyer}, {Bradley}, {Brady}, {Braun}, {Breda}, {Bresnahan}, {Bright}, {Britt}, {Bromenschenkel}, {Brooks}, {Brooks}, {Brown}, {Brown}, {Brown}, {Bunker}, {Burger}, {Bushouse}, {Cale}, {Cameron},
  {Cameron}, {Canipe}, {Caplinger}, {Caputo}, {Cara}, {Carey}, {Carniani}, {Carrasquilla}, {Carruthers}, {Case}, {Catherine}, {Chance}, {Chapman}, {Charlot}, {Charlow}, {Chayer}, {Chen}, {Cherinka}, {Chichester}, {Chilton}, {Chonis}, {Clampin}, {Clark}, {Clark}, {Coe}, {Coleman}, {Comber}, {Comeau}, {Connolly}, {Cooper}, {Cooper}, {Coppock}, {Correnti}, {Cossou}, {Coulais}, {Coyle}, {Cracraft}, {Curti}, {Cuturic}, {Davis}, {Davis}, {Dean}, {DeLisa}, {deMeester}, {Dencheva}, {Dencheva}, {DePasquale}, {Deschenes}, {Hunor Detre}, {Diaz}, {Dicken}, {DiFelice}, {Dillman}, {Dixon}, {Doggett}, {Donaldson}, {Douglas}, {DuPrie}, {Dupuis}, {Durning}, {Easmin}, {Eck}, {Edeani}, {Egami}, {Ehrenwinkler}, {Eisenhamer}, {Eisenhower}, {Elie}, {Elliott}, {Elliott}, {Ellis}, {Engesser}, {Espinoza}, {Etienne}, {Etxaluze}, {Falini}, {Feeney}, {Ferry}, {Filippazzo}, {Fincham}, {Fix}, {Flagey}, {Florian}, {Flynn}, {Fontanella}, {Ford}, {Forshay}, {Fox}, {Franz}, {Fu}, {Fullerton}, {Galkin}, {Galyer}, {Garc{\'\i}a Mar{\'\i}n},
  {Gardner}, {Gardner}, {Garland}, {Garrett}, {Gasman}, {Gaspar}, {Gaudreau}, {Gauthier}, {Geers}, {Geithner}, {Gennaro}, {Giardino}, {Girard}, {Giuliano}, {Glassmire}, {Glauser}, {Glazer}, {Godfrey}, {Golimowski}, {Gollnitz}, {Gong}, {Gonzaga}, {Gordon}, {Gordon}, {Goudfrooij}, {Greene}, {Greenhouse}, {Grimaldi}, {Groebner}, {Grundy}, {Guillard}, {Gutman}, {Ha}, {Haderlein}, {Hagedorn}, {Hainline}, {Haley}, {Hami}, {Hamilton}, {Hammel}, {Hansen}, {Harkins}, {Harr}, {Hart}, {Hart}, {Hartig}, {Hashimoto}, {Haskins}, {Hathaway}, {Havey}, {Hayden}, {Hecht}, {Heller-Boyer}, {Henriques}, {Henry}, {Hermann}, {Hernandez}, {Hesman}, {Hicks}, {Hilbert}, {Hines}, {Hoffman}, {Holfeltz}, {Holler}, {Hoppa}, {Hott}, {Howard}, {Howard}, {Hunter}, {Hunter}, {Hurst}, {Husemann}, {Hustak}, {Ilinca Ignat}, {Illingworth}, {Irish}, {Jackson}, {Jahromi}, {Jakobsen}, {James}, {James}, {Januszewski}, {Jenkins}, {Jirdeh}, {Johnson}, {Johnson}, {Jones}, {Jones}, {Jones}, {Jones}, {Jordan}, {Jordan}, {Jurczyk}, {Jurling}, {Kaleida},
  {Kalmanson}, {Kammerer}, {Kang}, {Kao}, {Karakla}, {Kavanagh}, {Kelly}, {Kendrew}, {Kennedy}, {Kenny}, {Keski-kuha}, {Keyes}, {Kidwell}, {Kinzel}, {Kirk}, {Kirkpatrick}, {Kirshenblat}, {Klaassen}, {Knapp}, {Knight}, {Knollenberg}, {Koehler}, {Koekemoer}, {Kovacs}, {Kulp}, {Kumari}, {Kyprianou}, {La Massa}, {Labador}, {Labiano}, {Lagage}, {Lajoie}, {Lallo}, {Lam}, {Lamb}, {Lambros}, {Lampenfield}, {Langston}, {Larson}, {Law}, {Lawrence}, {Lee}, {Leisenring}, {Lepo}, {Leveille}, {Levenson}, {Levine}, {Levy}, {Lewis}, {Lewis}, {Libralato}, {Lightsey}, {Link}, {Liu}, {Lo}, {Lockwood}, {Logue}, {Long}, {Long}, {Loomis}, {Lopez-Caniego}, {Lorenzo Alvarez}, {Love-Pruitt}, {Lucy}, {Luetzgendorf}, {Maghami}, {Maiolino}, {Major}, {Malla}, {Malumuth}, {Manjavacas}, {Mannfolk}, {Marrione}, {Marston}, {Martel}, {Maschmann}, {Masci}, {Masciarelli}, {Maszkiewicz}, {Mather}, {McKenzie}, {McLean}, {McMaster}, {Melbourne}, {Mel{\'e}ndez}, {Menzel}, {Merz}, {Meyett}, {Meza}, {Miskey}, {Misselt}, {Moller}, {Morrison}, {Morse},
  {Moseley}, {Mosier}, {Mountain}, {Mueckay}, {Mueller}, {Mullally}, {Murphy}, {Murray}, {Murray}, {Mustelier}, {Muzerolle}, {Mycroft}, {Myers}, {Myrick}, {Nanavati}, {Nance}, {Nayak}, {Naylor}, {Nelan}, {Nickson}, {Nielson}, {Nieto-Santisteban}, {Nikolov}, {Noriega-Crespo}, {O'Shaughnessy}, {O'Sullivan}, {Ochs}, {Ogle}, {Oleszczuk}, {Olmsted}, {Osborne}, {Ottens}, {Owens}, {Pacifici}, {Pagan}, {Page}, {Park}, {Parrish}, {Patapis}, {Paul}, {Pauly}, {Pavlovsky}, {Pedder}, {Peek}, {Pena-Guerrero}, {Penanen}, {Perez}, {Perna}, {Perriello}, {Phillips}, {Pietraszkiewicz}, {Pinaud}, {Pirzkal}, {Pitman}, {Piwowar}, {Platais}, {Player}, {Plesha}, {Pollizi}, {Polster}, {Pontoppidan}, {Porterfield}, {Proffitt}, {Pueyo}, {Pulliam}, {Quirt}, {Quispe Neira}, {Ramos Alarcon}, {Ramsay}, {Rapp}, {Rapp}, {Rauscher}, {Ravindranath}, {Rawle}, {Regan}, {Reichard}, {Reis}, {Ressler}, {Rest}, {Reynolds}, {Rhue}, {Richon}, {Rickman}, {Ridgaway}, {Ritchie}, {Rix}, {Robberto}, {Robinson}, {Robinson}, {Robinson}, {Rock}, {Rodriguez},
  {Rodriguez Del Pino}, {Roellig}, {Rohrbach}, {Roman}, {Romelfanger}, {Rose}, {Roteliuk}, {Roth}, {Rothwell}, {Rowlands}, {Roy}, {Royer}, {Royle}, {Rui}, {Rumler}, {Runnels}, {Russ}, {Rustamkulov}, {Ryden}, {Ryer}, {Sabata}, {Sabatke}, {Sabbi}, {Samuelson}, {Sapp}, {Sappington}, {Sargent}, {Sauer}, {Scheithauer}, {Schlawin}, {Schlitz}, {Schmitz}, {Schneider}, {Schreiber}, {Schulze}, {Schwab}, {Scott}, {Sembach}, {Shanahan}, {Shaughnessy}, {Shaw}, {Shawger}, {Shay}, {Sheehan}, {Shen}, {Sherman}, {Shiao}, {Shih}, {Shivaei}, {Sienkiewicz}, {Sing}, {Sirianni}, {Sivaramakrishnan}, {Skipper}, {Sloan}, {Slocum}, {Slowinski}, {Smith}, {Smith}, {Smith}, {Smith}, {Snyder}, {Soh}, {Sohn}, {Soto}, {Spencer}, {Stallcup}, {Stansberry}, {Starr}, {Starr}, {Stewart}, {Stiavelli}, {Straughn}, {Strickland}, {Stys}, {Summers}, {Sun}, {Sunnquist}, {Swade}, {Swam}, {Swaters}, {Swoish}, {Taylor}, {Taylor}, {Te Plate}, {Tea}, {Teague}, {Telfer}, {Temim}, {Thatte}, {Thompson}, {Thompson}, {Thomson}, {Tikkanen}, {Tippet}, {Todd},
  {Toolan}, {Tran}, {Trejo}, {Truong}, {Tsukamoto}, {Tustain}, {Tyra}, {Ubeda}, {Underwood}, {Uzzo}, {Van Campen}, {Vandal}, {Vandenbussche}, {Vila}, {Volk}, {Wahlgren}, {Waldman}, {Walker}, {Wander}, {Warfield}, {Warner}, {Wasiak}, {Watkins}, {Weaver}, {Weilert}, {Weiser}, {Weiss}, {Weissman}, {Welty}, {West}, {Wheate}, {Wheatley}, {Wheeler}, {White}, {Whiteaker}, {Whitehouse}, {Whiteleather}, {Whitman}, {Williams}, {Willmer}, {Willoughby}, {Wilson}, {Wirth}, {Wislowski}, {Wolf}, {Wolfe}, {Wolff}, {Workman}, {Wright}, {Wu}, {Wu}, {Wymer}, {Yates}, {Yeager}, {Yeates}, {Yerger}, {Yoon}, {Young}, {Yu}, {Zak}, {Zeidler}, {Zhou}, {Zielinski}, {Zincke}, \& {Zonak}}]{Rigby2023}
{Rigby}, J., {Perrin}, M., {McElwain}, M., {et~al.} 2023{\natexlab{a}}, \pasp, 135, 048001

\bibitem[{{Rigby} {et~al.}(2023{\natexlab{b}}){Rigby}, {Vieira}, {Phadke}, {Hutchison}, {Welch}, {Cathey}, {Spilker}, {Gonzalez}, {Adhikari}, {Aravena}, {Bayliss}, {Birkin}, {Bursk}, {Chapman}, {Dahle}, {Elicker}, {Fischer}, {Florian}, {Gladders}, {Hayward}, {Hewald}, {Kettler}, {Khullar}, {Kim}, {Law}, {Mahler}, {Malhotra}, {Murphy}, {Narayanan}, {Olivier}, {Rhoads}, {Sharon}, {Solimano}, {Thiruvengadam}, {Vizgan}, \& {Younker}}]{Rigby2023b}
{Rigby}, J.~R., {Vieira}, J.~D., {Phadke}, K.~A., {et~al.} 2023{\natexlab{b}}, \apj~in review, arXiv:2312.10465

\bibitem[{{Rodr{\'\i}guez Del Pino} {et~al.}(2024){Rodr{\'\i}guez Del Pino}, {Perna}, {Arribas}, {D'Eugenio}, {Lamperti}, {P{\'e}rez-Gonz{\'a}lez}, {{\"U}bler}, {Bunker}, {Carniani}, {Charlot}, {Maiolino}, {Willott}, {B{\"o}ker}, {Chevallard}, {Cresci}, {Curti}, {Jones}, {Parlanti}, {Scholtz}, \& {Venturi}}]{Pino2023}
{Rodr{\'\i}guez Del Pino}, B., {Perna}, M., {Arribas}, S., {et~al.} 2024, \aap, 684, A187

\bibitem[{Rodr{\'{i}}guez-Mu{\~{n}}oz {et~al.}(2021)Rodr{\'{i}}guez-Mu{\~{n}}oz, Rodighiero, P{\'{e}}rez-Gonz{\'{a}}lez, Talia, Baronchelli, Morselli, Renzini, Puglisi, Grazian, Zanella, Mancini, Feltre, Romano, {Vidal Garc{\'{i}}a}, Franceschini, {Alcalde Pampliega}, Cassata, Costantin, {Dom{\'{i}}nguez S{\'{a}}nchez}, Espino-Briones, Iani, Koekemoer, Lumbreras-Calle, \& Rodr{\'{i}}guez-Espinosa}]{Rodriguez-Munoz2021}
Rodr{\'{i}}guez-Mu{\~{n}}oz, L., Rodighiero, G., P{\'{e}}rez-Gonz{\'{a}}lez, P.~G., {et~al.} 2021, MNRAS, 510, 2061

\bibitem[{Ryder \& Dopita(1994)}]{Ryder1994}
Ryder, S.~D. \& Dopita, M.~A. 1994, ApJ, 430, 142

\bibitem[{{S. Gonzaga, W. Hack, A. Fruchter}(2012)}]{Gonzaga2012}
{S. Gonzaga, W. Hack, A. Fruchter}, J.~M. 2012, STScI, 63

\bibitem[{{Saxena} {et~al.}(2024){Saxena}, {Overzier}, {Villar-Mart{\'\i}n}, {Heckman}, {Roy}, {Duncan}, {R{\"o}ttgering}, {Miley}, {Aydar}, {Best}, {Bosman}, {Cameron}, {Gab{\'a}nyi}, {Humphrey}, {Morais}, {Onoue}, {Pentericci}, {Reynaldi}, \& {Venemans}}]{Saxena2024}
{Saxena}, A., {Overzier}, R.~A., {Villar-Mart{\'\i}n}, M., {et~al.} 2024, \mnras, 531, 4391

\bibitem[{{Shen} {et~al.}(2024){Shen}, {Papovich}, {Matharu}, {Pirzkal}, {Hu}, {Backhaus}, {Bagley}, {Cheng}, {Cleri}, {Finkelstein}, {Huertas-Company}, {Giavalisco}, {Grogin}, {Jung}, {Kartaltepe}, {Koekemoer}, {Lotz}, {Maseda}, {P{\'e}rez-Gonz{\'a}lez}, {Rothberg}, {Simons}, {Tacchella}, {Williams}, \& {Yung}}]{Shen2023b}
{Shen}, L., {Papovich}, C., {Matharu}, J., {et~al.} 2024, \apjl, 963, L49

\bibitem[{{Shen} {et~al.}(2023){Shen}, {Papovich}, {Yang}, {Matharu}, {Wang}, {Magnelli}, {Elbaz}, {Jogee}, {Alavi}, {Arrabal Haro}, {Backhaus}, {Bagley}, {Bell}, {Bisigello}, {Calabr{\`o}}, {Cooper}, {Costantin}, {Daddi}, {Dickinson}, {Finkelstein}, {Fujimoto}, {Giavalisco}, {Grogin}, {Guo}, {Holwerda}, {Kartaltepe}, {Koekemoer}, {Kurczynski}, {Lucas}, {P{\'e}rez-Gonz{\'a}lez}, {Pirzkal}, {Prichard}, {Rafelski}, {Ronayne}, {Simons}, {Sunnquist}, {Teplitz}, {Trump}, {Weiner}, {Windhorst}, \& {Yung}}]{Shen2023a}
{Shen}, L., {Papovich}, C., {Yang}, G., {et~al.} 2023, \apj, 950, 7

\bibitem[{{Shivaei} {et~al.}(2020){Shivaei}, {Reddy}, {Rieke}, {Shapley}, {Kriek}, {Battisti}, {Mobasher}, {Sanders}, {Fetherolf}, {Azadi}, {Coil}, {Freeman}, {de Groot}, {Leung}, {Price}, {Siana}, \& {Zick}}]{Shivaei2020}
{Shivaei}, I., {Reddy}, N., {Rieke}, G., {et~al.} 2020, \apj, 899, 117

\bibitem[{{Simons} {et~al.}(2017){Simons}, {Kassin}, {Weiner}, {Faber}, {Trump}, {Heckman}, {Koo}, {Pacifici}, {Primack}, {Snyder}, \& {de la Vega}}]{Simons2017}
{Simons}, R.~C., {Kassin}, S.~A., {Weiner}, B.~J., {et~al.} 2017, \apj, 843, 46

\bibitem[{{Simons} {et~al.}(2021){Simons}, {Papovich}, {Momcheva}, {Trump}, {Brammer}, {Estrada-Carpenter}, {Backhaus}, {Cleri}, {Finkelstein}, {Giavalisco}, {Ji}, {Jung}, {Matharu}, \& {Weiner}}]{Simons2020a}
{Simons}, R.~C., {Papovich}, C., {Momcheva}, I., {et~al.} 2021, \apj, 923, 203

\bibitem[{{Simons} {et~al.}(2023){Simons}, {Papovich}, {Momcheva}, {Brammer}, {Estrada-Carpenter}, {Finkelstein}, {Gosmeyer}, {Matharu}, {Trump}, {Backhaus}, {Cheng}, {Cleri}, {Ferguson}, {Finlator}, {Giavalisco}, {Ji}, {Jung}, {Lotz}, {O'Brien}, {Skelton}, {Tilvi}, \& {Weiner}}]{Simons2023}
{Simons}, R.~C., {Papovich}, C., {Momcheva}, I.~G., {et~al.} 2023, \apjs, 266, 13

\bibitem[{{Str{\"o}mgren}(1939)}]{Stromgren1939}
{Str{\"o}mgren}, B. 1939, \apj, 89, 526

\bibitem[{{Tacconi} {et~al.}(2013){Tacconi}, {Neri}, {Genzel}, {Combes}, {Bolatto}, {Cooper}, {Wuyts}, {Bournaud}, {Burkert}, {Comerford}, {Cox}, {Davis}, {F{\"o}rster Schreiber}, {Garc{\'\i}a-Burillo}, {Gracia-Carpio}, {Lutz}, {Naab}, {Newman}, {Omont}, {Saintonge}, {Shapiro Griffin}, {Shapley}, {Sternberg}, \& {Weiner}}]{Tacconi2013}
{Tacconi}, L.~J., {Neri}, R., {Genzel}, R., {et~al.} 2013, \apj, 768, 74

\bibitem[{Theios {et~al.}(2019)Theios, Steidel, Strom, Rudie, Trainor, \& Reddy}]{Theios2019}
Theios, R.~L., Steidel, C.~C., Strom, A.~L., {et~al.} 2019, ApJ, 871, 128

\bibitem[{{Turner} {et~al.}(2017){Turner}, {Cirasuolo}, {Harrison}, {McLure}, {Dunlop}, {Swinbank}, {Johnson}, {Sobral}, {Matthee}, \& {Sharples}}]{Turner2017}
{Turner}, O.~J., {Cirasuolo}, M., {Harrison}, C.~M., {et~al.} 2017, \mnras, 471, 1280

\bibitem[{{{\"U}bler} {et~al.}(2019){{\"U}bler}, {Genzel}, {Wisnioski}, {F{\"o}rster Schreiber}, {Shimizu}, {Price}, {Tacconi}, {Belli}, {Wilman}, {Fossati}, {Mendel}, {Davies}, {Beifiori}, {Bender}, {Brammer}, {Burkert}, {Chan}, {Davies}, {Fabricius}, {Galametz}, {Herrera-Camus}, {Lang}, {Lutz}, {Momcheva}, {Naab}, {Nelson}, {Saglia}, {Tadaki}, {van Dokkum}, \& {Wuyts}}]{Ubler2019}
{{\"U}bler}, H., {Genzel}, R., {Wisnioski}, E., {et~al.} 2019, \apj, 880, 48

\bibitem[{{{\"U}bler} {et~al.}(2023){{\"U}bler}, {Maiolino}, {Curtis-Lake}, {P{\'e}rez-Gonz{\'a}lez}, {Curti}, {Perna}, {Arribas}, {Charlot}, {Marshall}, {D'Eugenio}, {Scholtz}, {Bunker}, {Carniani}, {Ferruit}, {Jakobsen}, {Rix}, {Rodr{\'\i}guez Del Pino}, {Willott}, {Boeker}, {Cresci}, {Jones}, {Kumari}, \& {Rawle}}]{Ubler2023}
{{\"U}bler}, H., {Maiolino}, R., {Curtis-Lake}, E., {et~al.} 2023, \aap, 677, A145

\bibitem[{{{\"U}bler} {et~al.}(2024){{\"U}bler}, {Maiolino}, {P{\'e}rez-Gonz{\'a}lez}, {D'Eugenio}, {Perna}, {Curti}, {Arribas}, {Bunker}, {Carniani}, {Charlot}, {Rodr{\'\i}guez Del Pino}, {Baker}, {B{\"o}ker}, {Cresci}, {Dunlop}, {Grogin}, {Jones}, {Kumari}, {Lamperti}, {Laporte}, {Marshall}, {Mazzolari}, {Parlanti}, {Rawle}, {Scholtz}, {Venturi}, \& {Witstok}}]{Ubler2023b}
{{\"U}bler}, H., {Maiolino}, R., {P{\'e}rez-Gonz{\'a}lez}, P.~G., {et~al.} 2024, \mnras, 531, 355

\bibitem[{{Van Den Bosch}(2001)}]{VanDenBosch2001}
{Van Den Bosch}, F.~C. 2001, MNRAS, 327, 1334

\bibitem[{van~der Wel {et~al.}(2014)van~der Wel, Franx, van Dokkum, Skelton, Momcheva, Whitaker, Brammer, Bell, Rix, Wuyts, Ferguson, Holden, Barro, Koekemoer, Chang, McGrath, H{\"{a}}ussler, Dekel, Behroozi, Fumagalli, Leja, Lundgren, Maseda, Nelson, Wake, Patel, Labb{\'{e}}, Faber, Grogin, \& Kocevski}]{VanderWel2014}
van~der Wel, A., Franx, M., van Dokkum, P.~G., {et~al.} 2014, ApJ, 788, 28

\bibitem[{van Dokkum {et~al.}(2011)van Dokkum, Brammer, Fumagalli, Nelson, Franx, Rix, Kriek, Skelton, Patel, Schmidt, Bezanson, Bian, da~Cunha, Erb, Fan, {F{\"{o}}rster Schreiber}, Illingworth, Labb{\'{e}}, Lundgren, Magee, Marchesini, McCarthy, Muzzin, Quadri, Steidel, Tal, Wake, Whitaker, \& Williams}]{VanDokkum2011}
van Dokkum, P.~G., Brammer, G., Fumagalli, M., {et~al.} 2011, ApJ, 743, L15

\bibitem[{Vollmer {et~al.}(2012)Vollmer, Soida, Braine, Abramson, Beck, Chung, Crowl, Kenney, \& {Van Gorkom}}]{Vollmer2012}
Vollmer, B., Soida, M., Braine, J., {et~al.} 2012, A{\&}A, 537, 1

\bibitem[{{Wang} {et~al.}(2024){Wang}, {Wylezalek}, {De Breuck}, {Vernet}, {Rupke}, {Zakamska}, {Vayner}, {Lehnert}, {Nesvadba}, \& {Stern}}]{Wang2024}
{Wang}, W., {Wylezalek}, D., {De Breuck}, C., {et~al.} 2024, \aap, 683, A169

\bibitem[{{Ward} {et~al.}(2024){Ward}, {de la Vega}, {Mobasher}, {McGrath}, {Iyer}, {Calabr{\`o}}, {Costantin}, {Dickinson}, {Holwerda}, {Huertas-Company}, {Hirschmann}, {Lucas}, {Pandya}, {Wilkins}, {Yung}, {Arrabal Haro}, {Bagley}, {Finkelstein}, {Kartaltepe}, {Koekemoer}, {Papovich}, \& {Pirzkal}}]{Ward2023}
{Ward}, E., {de la Vega}, A., {Mobasher}, B., {et~al.} 2024, \apj, 962, 176

\bibitem[{{Weibel} {et~al.}(2024){Weibel}, {Oesch}, {Barrufet}, {Gottumukkala}, {Ellis}, {Santini}, {Weaver}, {Allen}, {Bouwens}, {Bowler}, {Brammer}, {Carnall}, {Cullen}, {Dayal}, {Donnan}, {Dunlop}, {Giavalisco}, {Grogin}, {Illingworth}, {Koekemoer}, {Labbe}, {Marchesini}, {McLeod}, {McLure}, {Naidu}, {Shuntov}, {Stefanon}, {Toft}, \& {Xiao}}]{Weibel2024}
{Weibel}, A., {Oesch}, P.~A., {Barrufet}, L., {et~al.} 2024, \mnras~in review, arXiv:2403.08872

\bibitem[{White \& Rees(1978)}]{White&Rees}
White, S. \& Rees, M. 1978, 183, 341

\bibitem[{{Willott} {et~al.}(2022){Willott}, {Doyon}, {Albert}, {Brammer}, {Dixon}, {Muzic}, {Ravindranath}, {Scholz}, {Abraham}, {Artigau}, {Brada{\v{c}}}, {Goudfrooij}, {Hutchings}, {Iyer}, {Jayawardhana}, {LaMassa}, {Martis}, {Meyer}, {Morishita}, {Mowla}, {Muzzin}, {Noirot}, {Pacifici}, {Rowlands}, {Sarrouh}, {Sawicki}, {Taylor}, {Volk}, \& {Zabl}}]{Willott2022}
{Willott}, C.~J., {Doyon}, R., {Albert}, L., {et~al.} 2022, \pasp, 134, 025002

\bibitem[{Wilman {et~al.}(2020)Wilman, Fossati, Mendel, Saglia, Wisnioski, Wuyts, Schreiber, Beifiori, Bender, Belli, {\"{U}}bler, Lang, Chan, Davies, Nelson, Genzel, Tacconi, Galametz, Davies, Lutz, Price, Burkert, Tadaki, Herrera-Camus, Brammer, Momcheva, \& van Dokkum}]{Wilman2020}
Wilman, D.~J., Fossati, M., Mendel, J.~T., {et~al.} 2020, ApJ, 892, 1

\bibitem[{{Wisnioski} {et~al.}(2019){Wisnioski}, {F{\"o}rster Schreiber}, {Fossati}, {Mendel}, {Wilman}, {Genzel}, {Bender}, {Wuyts}, {Davies}, {{\"U}bler}, {Bandara}, {Beifiori}, {Belli}, {Brammer}, {Chan}, {Davies}, {Fabricius}, {Galametz}, {Lang}, {Lutz}, {Nelson}, {Momcheva}, {Price}, {Rosario}, {Saglia}, {Seitz}, {Shimizu}, {Tacconi}, {Tadaki}, {van Dokkum}, \& {Wuyts}}]{Wisnioski2019}
{Wisnioski}, E., {F{\"o}rster Schreiber}, N.~M., {Fossati}, M., {et~al.} 2019, \apj, 886, 124

\bibitem[{Wisnioski {et~al.}(2015)Wisnioski, {F{\"{o}}rster Schreiber}, Wuyts, Wuyts, Bandara, Wilman, Genzel, Bender, Davies, Fossati, Lang, Mendel, Beifiori, Brammer, Chan, Fabricius, Fudamoto, Kulkarni, Kurk, Lutz, Nelson, Momcheva, Rosario, Saglia, Seitz, Tacconi, \& van Dokkum}]{Wisnioski2015}
Wisnioski, E., {F{\"{o}}rster Schreiber}, N.~M., Wuyts, S., {et~al.} 2015, ApJ, 799, 209

\bibitem[{{Wright} {et~al.}(2023){Wright}, {Rieke}, {Glasse}, {Ressler}, {Garc{\'\i}a Mar{\'\i}n}, {Aguilar}, {Alberts}, {{\'A}lvarez-M{\'a}rquez}, {Argyriou}, {Banks}, {Baudoz}, {Boccaletti}, {Bouchet}, {Bouwman}, {Brandl}, {Breda}, {Bright}, {Cale}, {Colina}, {Cossou}, {Coulais}, {Cracraft}, {De Meester}, {Dicken}, {Engesser}, {Etxaluze}, {Fox}, {Friedman}, {Fu}, {Gasman}, {G{\'a}sp{\'a}r}, {Gastaud}, {Geers}, {Glauser}, {Gordon}, {Greene}, {Greve}, {Grundy}, {G{\"u}del}, {Guillard}, {Haderlein}, {Hashimoto}, {Henning}, {Hines}, {Holler}, {Detre}, {Jahromi}, {James}, {Jones}, {Justtanont}, {Kavanagh}, {Kendrew}, {Klaassen}, {Krause}, {Labiano}, {Lagage}, {Lambros}, {Larson}, {Law}, {Lee}, {Libralato}, {Lorenzo Alverez}, {Meixner}, {Morrison}, {Mueller}, {Murray}, {Mycroft}, {Myers}, {Nayak}, {Naylor}, {Nickson}, {Noriega-Crespo}, {{\"O}stlin}, {O'Sullivan}, {Ottens}, {Patapis}, {Penanen}, {Pietraszkiewicz}, {Ray}, {Regan}, {Roteliuk}, {Royer}, {Samara-Ratna}, {Samuelson}, {Sargent}, {Scheithauer},
  {Schneider}, {Schreiber}, {Shaughnessy}, {Sheehan}, {Shivaei}, {Sloan}, {Tamas}, {Teague}, {Temim}, {Tikkanen}, {Tustain}, {van Dishoeck}, {Vandenbussche}, {Weilert}, {Whitehouse}, \& {Wolff}}]{Wright2023}
{Wright}, G.~S., {Rieke}, G.~H., {Glasse}, A., {et~al.} 2023, \pasp, 135, 048003

\bibitem[{Wuyts {et~al.}(2011)Wuyts, {F{\"{o}}rster Schreiber}, Lutz, Nordon, Berta, Altieri, Andreani, Aussel, Bongiovanni, Cepa, Cimatti, Daddi, Elbaz, Genzel, Koekemoer, Magnelli, Maiolino, McGrath, Garc{\'{i}}a, Poglitsch, Popesso, Pozzi, Sanchez-Portal, Sturm, Tacconi, \& Valtchanov}]{Wuyts2011}
Wuyts, S., {F{\"{o}}rster Schreiber}, N.~M., Lutz, D., {et~al.} 2011, ApJ, 738, 106

\bibitem[{{Wuyts} {et~al.}(2013){Wuyts}, {F{\"o}rster Schreiber}, {Nelson}, {van Dokkum}, {Brammer}, {Chang}, {Faber}, {Ferguson}, {Franx}, {Fumagalli}, {Genzel}, {Grogin}, {Kocevski}, {Koekemoer}, {Lundgren}, {Lutz}, {McGrath}, {Momcheva}, {Rosario}, {Skelton}, {Tacconi}, {van der Wel}, \& {Whitaker}}]{Wuyts2013}
{Wuyts}, S., {F{\"o}rster Schreiber}, N.~M., {Nelson}, E.~J., {et~al.} 2013, \apj, 779, 135

\bibitem[{Yoshikawa {et~al.}(2010)Yoshikawa, Akiyama, Kajisawa, Alexander, Ohta, Suzuki, Tokoku, Uchimoto, Konishi, Yamada, Tanaka, Omata, Nishimura, Koekemoer, Brandt, \& Ichikawa}]{Yoshikawa2010}
Yoshikawa, T., Akiyama, M., Kajisawa, M., {et~al.} 2010, ApJ, 718, 112

\end{thebibliography}
\bibliographystyle{aa}

\end{document}